\documentclass[modern]{aastex63Rev}
\usepackage{amssymb, amsmath}
\usepackage{color}
\usepackage{graphicx}
\usepackage{morefloats}
\usepackage{hyperref}
\NewPageAfterKeywords

\begin{document}

\title{The Time-Averaged Mass-Loss Rates of Red Supergiants As Revealed by their Luminosity Functions in M31 and M33}

\author[0000-0001-6563-7828]{Philip Massey}
\affiliation{Lowell Observatory, 1400 W Mars Hill Road, Flagstaff, AZ 86001, USA}
\affiliation{Department of Astronomy and Planetary Science, Northern Arizona University, Flagstaff, AZ, 86011-6010, USA}
\email{phil.massey@lowell.edu}

\author[0000-0002-5787-138X]{Kathryn F. Neugent}
\altaffiliation{NASA Hubble Fellow}
\affiliation{Center for Astrophysics, Harvard \& Smithsonian, 60 Garden St., Cambridge, MA 02138, USA}
\affiliation{Lowell Observatory, 1400 W Mars Hill Road, Flagstaff, AZ 86001, USA}

\author[0000-0002-2564-5660]{Sylvia Ekstr\"{o}m}
\affiliation{Geneva Observatory, University of Geneva, Chemin Pegasi 51, 1290, Sauverny, Switzerland}

\author[0000-0003-2362-4089]{Cyril Georgy}
\affiliation{Geneva Observatory, University of Geneva, Chemin Pegasi 51, 1290, Sauverny, Switzerland}

\author[0000-0001-6181-1323]{Georges Meynet}
\affiliation{Geneva Observatory, University of Geneva, Chemin Pegasi 51, 1290, Sauverny, Switzerland}

\begin{abstract}

Mass-loss in red supergiants (RSGs) is generally recognized to be episodic, but mass-loss prescriptions fail to reflect this. Evolutionary models show that the total amount of mass lost during this phase determines if these stars evolve to warmer temperatures
before undergoing core collapse.  The current Geneva evolutionary models mimic episodic mass loss by enhancing the quiescent prescription rates whenever the star's outer layers exceed the Eddington luminosity by a large factor.  This results in a 20$M_\odot$ model undergoing significantly more mass loss during the RSG phase than it would have otherwise, but has little effect on models of lower masses. We can test the validity of this approach observationally by measuring the proportion of high-luminosity RSGs to that predicted by the models.  To do this, we use our recent luminosity-limited census of RSGs in M31 and M33, making modest improvements to membership, and adopting extinctions based on the recent panchromatic M31 and M33 Hubble surveys.  We then compare the proportions of the highest luminosity RSGs found to that predicted by published Geneva models, as well as to a special set of models computed without the enhanced rates.  We find good agreement with the models which include the supra-Eddington enhanced mass loss. The models with lower mass-loss rates predict a larger fraction of high-luminosity RSGs than observed, and thus can be ruled out. We also use these improved data to confirm that the upper luminosity limit of RSGs is $\log L/L_\odot\sim5.4$, regardless of metallicity, using our improved data on M31 and M33 plus previous results on the Magellanic Clouds.
\end{abstract}

\section{Introduction}

Understanding the evolution of massive stars is important not just in its own right, but also for solving a host of other problems of astrophysical interest, such as modeling the chemical evolution of galaxies, determining the origins of gamma-ray bursts, and interpreting the gravitational wave signatures of merging compact objects.   The treatment of convection, the inclusion of magnetic fields, and the distribution of the initial rotation rates all are uncertainties in the construction of massive star evolutionary models. However, \citet{2015EAS....71...41G} argue that one of the least constrained and most important properties is that of the time-averaged mass-loss rate during the red supergiant (RSG) phase. Such mass loss shortens the lifetime of the RSG phase, and, at high values, results in such stars shedding their hydrogen-rich outer layers and evolving back towards higher temperatures before ending their lives as supernovae (SNe).    It also changes the distribution of SN types expected, and possibly even the sort of compact objects left behind, depending upon how much mass is lost during earlier phases.

There is a strong theoretical underpinning to our knowledge of OB stellar winds, where radiation pressure on a multitude of subordinate spectral lines drives the winds (\citealt{1975ApJ...195..157C}; for reviews, see \citealt{2008A&ARv..16..209P}, \citealt{2009AIPC.1171..111H}, and \citealt{2017IAUS..329..215S}).  By contrast, there is little understanding of what drives the winds in RSGs.  It is often hypothesized that RSG winds are driven by radiation pressure on dust grains. However, there is no clear physical mechanism for lifting the gas above the stellar surface to radii where it is cool enough that dust can form (see, e.g., \citealt{2015IAUS..307...25P}).  Although pulsations are powerful enough to do this in carbon-rich asymptotic giant branch (AGB) stars \citep{2009ASPC..414....3H}, they likely are not in RSGs, and turbulence may play an important role, as argued by \citet{2007A&A...469..671J} and \citet{2021A&A...646A.180K}. \citet{2018ApJ...869....1R,2019ApJ...882...37R} has demonstrated the importance of magnetohydrodynamic Alfv\'{e}n waves in the chromospheres of AGBs in lifting the material;  this may also play a role in RSGs.  However,  what drives the winds in RSGs remains an unsolved problem at present (see, e.g., \citealt{2015IAUS..307..280W}).    

Thus it is not yet possible to calculate the mass-loss rates of RSGs from first principles.  Instead, we have long
relied purely on empirical prescriptions, such as those given in \citet{1975MSRSL...8..369R}, \citet{1988A&AS...72..259D}, and \citet{2005A&A...438..273V}.  In general these prescriptions provide a formula for the mass-loss rates as a function of luminosity and effective temperature.  However, the actual data shows a tremendous scatter (2-3 orders of magnitude) about these simple prescriptions (see, e.g., Figure 1 in \citealt{2015A&A...575A..60M}). 

One contributing factor to this scatter is that RSGs of different masses and surface gravities can have the same effective temperatures and luminosities:  RSGs evolve  nearly vertically along a Hayashi track in the Hertzsprung-Russell diagram (HRD), as the luminosity is a steep function of the mean particle mass, which increases as helium is converted into carbon and oxygen.  (See also the discussion in \citealt{ 2020MNRAS.494L..53F}.) \citet{2020MNRAS.492.5994B} attempted to get around this issue by measuring the RSG mass-loss rates in clusters; in each cluster, the range of RSG masses will be very small, assuming the cluster formed coevally.  \citet{2020MNRAS.492.5994B} offer their own mass-loss prescription as a function of initial mass and luminosity, with values that are far lower than that of other studies.

However, there is potentially a much larger problem than taking into account the masses.  All of these various mass-loss prescriptions are based upon the measured ``instantaneous" mass loss rates of RSGs, not values averaged over the lifetime of the RSG phase. This is of particular concern as RSG mass loss is known to be heavily episodic, particularly at higher luminosities. 
The \citet{1999A&A...351..559V} study suggests that LMC RSGs shed {\it most} of their outer layers in short bursts of intense mass-loss.  Thus, prescriptions derived from ``quiescent" periods tell us little about what is needed from an evolutionary point of view.
These periods of high mass loss are likely due to the outer layers of the star exceeding their Eddington luminosity.  As explained by \citet{2009pfer.book.....M} and \citet{Sylvia}, this unstable situation is created as a result of the high opacity due to variations in the hydrogen ionization beneath the surface of the star.   Although these phases have not been observed directly, they can leave evidence in the circumstellar envelopes (see, e.g., \citealt{2009AJ....137.3558S} and \citealt{2011MNRAS.415..199M}).  For instance, \citet{DecinVYCMa} has shown that the high luminosity RSG VY CMa underwent a short (100~yr) burst of high mass-loss some 1000 years ago.  The 2019/2020 ``great dimming" of Betelgeuse was likely caused by ``mini-burst" of dust formation \citep{2020ApJ...891L..37L,2021Natur.594..365M,2022ApJ...936...18D}, underscoring that mass-loss in RSGs has a stochastic nature even outside of the supra-Eddington phase.  (For a contrasting view that suggests episodic mass loss is not a common phenomenon among RSGs, see \citealt{2022ApJ...933...41B}.)

Without a full hydrodynamical treatment, there is no precise way to include these mass-loss bursts in evolution models, and one-dimensional models require some approximate way of incorporating both the quiescent and outburst phases.  To achieve this, \citet{Sylvia} (somewhat arbitrarily) adopted a mass-loss rate of $3\times$ that of the quiescent state whenever the calculated luminosity of the RSG envelope exceeds a value of $5\times$ the Eddington luminosity; these are referred to as the ``supra-Eddington" periods.  (Their quiescent prescription comes from \citealt{1975MSRSL...8..369R,1977A&A....61..217R} with $\eta=0.5$ for RSGs with masses up to 12$M_\odot$, and from their own fit to the data in Figure~3 of \citealt{2001ASSL..264..215C} for more massive RSGs; the latter is roughly a factor of three times larger than the \citealt{1988A&AS...72..259D} rate.) The result of this is a time-averaged mass-loss rate that is a factor of ten higher for a 20$M_\odot$ RSG than what would have earlier been assumed  using the one of the older prescriptions  \citep{Sylvia}.

\begin{figure}[t]
\begin{centering}
\includegraphics[width=0.7\textwidth]{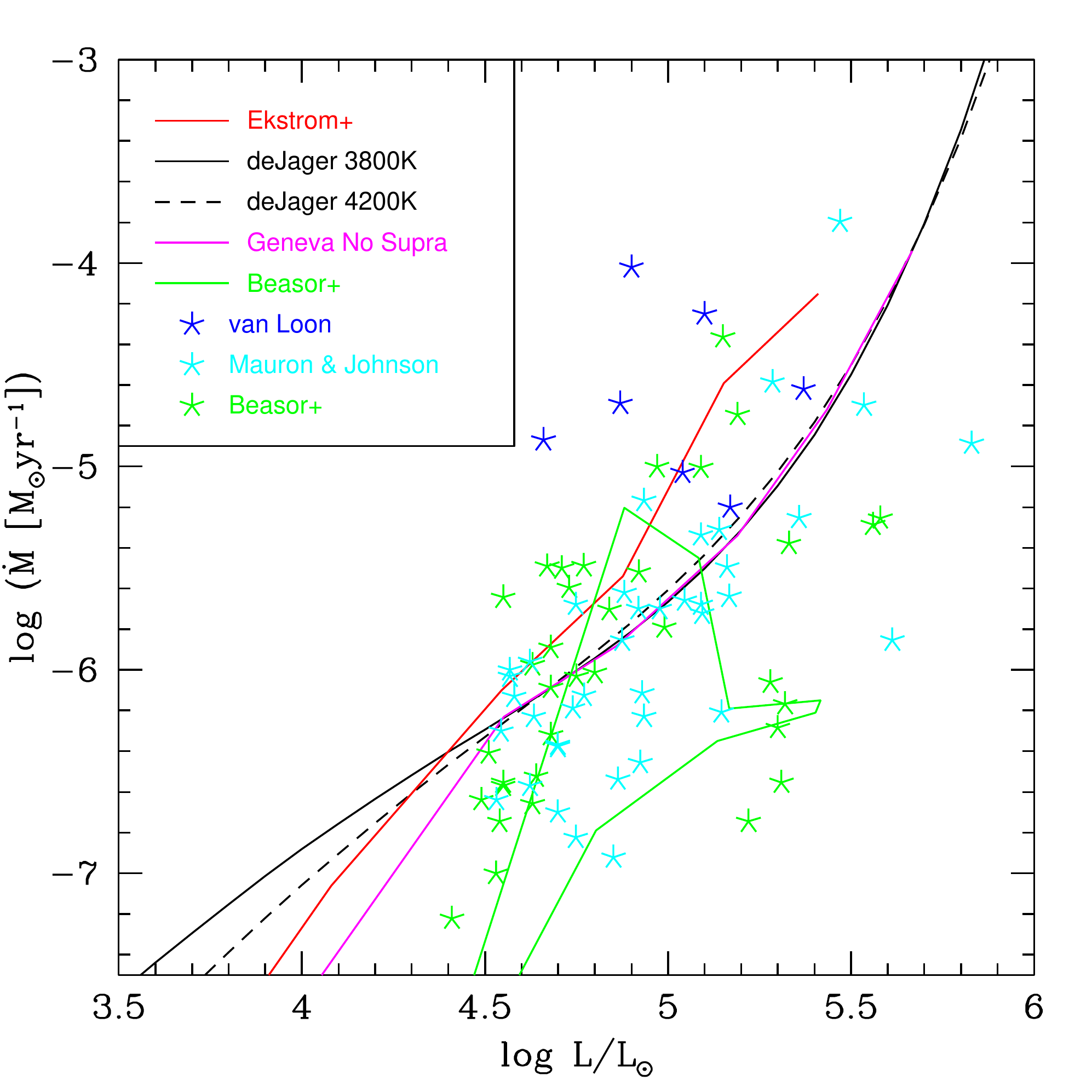}
\caption{\label{fig:ML} RSG mass-loss prescriptions compared to ``instantaneous" observed rates.  The red line shows the time-averaged mass-loss rates of RSGs in the models of \citet{Sylvia}, while the lower magenta line comes from the new set of models analyzed here that lack the supra-Eddington mass-loss rates.  The two black curves show the mass-loss prescription of \citet{1988A&AS...72..259D}, with the solid line corresponding to an effective temperature of 3800~K, while the dashed line corresponds to 4200~K.  The irregular region outlined in green corresponds to the region of mass-loss expected by the  new prescription offered by \citet{2020MNRAS.492.5994B} computed for the luminosity ranges covered by stars of 12, 15, 20, and 25$M_\odot$.  Blue points show the mass-loss rates for the subset of RSGs in the  \citet{2005A&A...438..273V} study.  Cyan points are from Table 1 in  \citet{2011A&A...526A.156M}.  Green points are from \citet{2018MNRAS.475...55B} and \citet{2020MNRAS.492.5994B}.} 
\end{centering}
\end{figure}

We show the effective Geneva time-averaged mass-loss rate as a function of luminosity by the red line in Figure~\ref{fig:ML}, compared to the classic \citet{1988A&AS...72..259D} prescription shown as the two black lines.
We emphasize the mass-loss rates are not
constant with the Geneva method because of the supra-Eddington phases.  To determine the time-averaged mass-loss rates for each mass track in the Geneva models,   we have simply taken the total mass lost during the red supergiant phase in the models and divided by the total time spent as a RSG.  Similarly the luminosity we have used is the time-averaged luminosity computed for the corresponding track.   We have also indicated the region of mass-loss rates that would be calculated from the \citet{2020MNRAS.492.5994B} mass-dependent relation, where we restrict the luminosity ranges to those from the \citet{Sylvia} models.  The figure includes various ``instantaneous" measurements of RSG mass-loss rates for comparison. The large range at a given
luminosity alluded to above is clear.  In adding these points we have excluded AGB stars, as it seems likely that their driving mechanisms are different.  (Our referee, Jacco van Loon, notes that the ``observed" mass-loss rates may be significantly underestimated for the RSGs with the lowest values, as these are the stars with inefficient dust formation.  Thus their actual dust-to-gas ratio is likely lower than what is usually assumed, and hence their mass-loss rates would be underestimated.) 

The  implications of including the supra-Eddington mass-loss in the evolutionary models are profound.  One of the most interesting implications is that it neatly solves the ``red supergiant problem" first noted by \citet{SmarttIIP}.  Type II-P supernovae have long been thought to originate from RSGs.  However, although RSGs are found with luminosities corresponding to 25-30$M_\odot$, \citet{SmarttIIP} found that the most luminous of the Type II-P progenitors correspond to masses of only 17-18$M_\odot$.  The \citet{Sylvia} calculations show that it is only RSGs above this limit that undergo supra-Eddington mass-loss, suggesting that these stars evolve back to the blue before undergoing core collapse, and hence likely exploding as SNe Type II-L or Ib, having shed most of their hydrogen-rich envelopes during these large mass-loss events\footnote{We note that, like most challenging discoveries, the red supergiant problem is not without its controversies. \citet{2018MNRAS.474.2116D}  argue that \citet{SmarttIIP} underestimated the bolometric corrections used to infer luminosities and hence masses, and that the red supergiant problem does not exist.  However, their approach has subsequently been criticized by \citet{2020MNRAS.493.4945K}, whose analysis comes to a similar conclusion as \citet{SmarttIIP}.}. 

\citet{2015A&A...575A..60M} suggested that the direct measurements of RSG mass-loss rates do not preclude their time-averaged values to be even higher, perhaps as much as an additional factor of 10-25$\times$.   In contrast, \citet{2018MNRAS.475...55B} and \citet{2020MNRAS.492.5994B} argue that these new evolutionary rates are already too high, although their arguments overlook the short periods of extremely high mass-loss that are expected as the outer layers exceed the Eddington luminosities. 

Fortunately, there is an observational test that we can use to address the issue.  \citet{2015A&A...575A..60M} and \citet{2015EAS....71...41G} note that the RSG luminosity function is quite sensitive to the total amount of mass lost during the RSG stage.  This is as expected, in that high mass loss rates will preferentially shorten the lifetimes of the highest luminosity RSGs, depopulating the top end of the RSG luminosity function.  Using a limited data set of RSG in M31, \cite{UKIRT} demonstrated that the luminosity function was consistent with the \citet{Sylvia} models that employed the enhanced mass-loss rates during the supra-Eddington phases, but inconsistent with the 10-25$\times$ further enhanced rates considered by \citet{2015A&A...575A..60M}.   

Here we reinvestigate the matter using more complete data on the RSG content of M31 and M33, including a comparison to
a newly computed set of Geneva models that are similar to those of \citet{Sylvia} in every way {\it except} that the mass-loss rates during the RSG phase have been lowered.  This has been done by ignoring any correction for the supra-Eddington phases and adopting the \citet{1988A&AS...72..259D} relation for the quiescent phase, rather than the slightly higher one derived from the \citet{2001ASSL..264..215C} data and used in the \citet{Sylvia} models.  The resulting time-averaged mass-loss rate is shown by the magenta line in Figure~\ref{fig:ML}; as expected, it follows the \citet{1988A&AS...72..259D} relation when $\log L/L_\odot>4.5$.  Although lower than the (now) standard Geneva rates, they are still much higher than those proposed by \citet{2020MNRAS.492.5994B}, particularly at larger luminosities and masses, as shown in Figure~\ref{fig:ML}.  The primary goal of our study is to see if the observed luminosity functions can successfully be used to rule out lower rates, or if these are, in fact, consistent with the observations.  In addition, our improvements to our derivation of the bolometric luminosities of these RSGs allow us to address the long-standing question of the upper luminosity limits of cool supergiants. In Section 2 we detail the construction of the luminosity functions, both theoretical and observational, and in Section 3 we compare the two, as well as discuss the upper luminosity limits of RSGs. Section 4 summarizes our work, and presents our conclusions.

\section{Constructing the Luminosity Functions}
\subsection{Theoretical}
\label{Sec-Theory}

In our preliminary study, \citet{UKIRT} compared the observed luminosity function of RSGs in M31 to that expected from three sets of Geneva evolutionary models: those computed with the normal mass-loss rates (including the supra-Eddington phases), plus those with 10$\times$ that rate and $25\times$ that rate.  Each set included evolutionary tracks that include rotation (initially 40\% of the breakup speed), and no rotation.  To these sets we now add the ``low mass-loss" models, computed similarly to those of \citet{Sylvia} but without the enhanced supra-Eddington mass-loss and with a pure \citet{1988A&AS...72..259D} prescription. All models were computed with solar metallicity $Z=0.014$. Further details can be found in \citet{Sylvia}.

Each set of evolutionary models tells us the expected effective temperature and luminosity at each time step for each of the individual mass tracks.   What we need, however, is the expected luminosity function for an ensemble of stars.  
To accomplish this we must carefully interpolate between the discrete mass tracks. In general, the mass tracks represented initial masses of 8$M_\odot$, 9$M_\odot$, 12$M_\odot$, 15$M_\odot$, 20$M_\odot$, 25$M_\odot$, and 32$M_\odot$; however, finer-grid tracks were created in the mass range between 8 - 15 $M_\odot$ to better model the Cepheid loops. 
Following, \citet{RSGWRs} and references therein, we assume a constant star-formation rate and a \citet{1955ApJ...121..161S} power-law slope of $\Gamma=-1.35$ ($\gamma=-2.35$) for the IMF, where the log of the number of stars
between masses $m_1$ and $m_2$ is simply proportional to $\Gamma \log m \Big|_{m_1}^{m_2}$ (see, e.g., \citealt{MasseyGilmore}).

This transformation was accomplished using the Geneva population synthesis code SYnthetic CLusters Isochrones \& Stellar Tracks ({\sc syclist}),  described by \citet{syclist}. 
We computed the luminosity functions using a minimum and maximum initial masses of 8$M_\odot$ and 30$M_\odot$, and restricted the sample to times when the stars had effective temperatures less than or equal to 4200~K.  We used 100,000 mass cells (``beams") in order to avoid quantization issues, and computed locations of stars in the HRD for single-aged populations with ages between  0 and $1.6 \times 10^8$ yr with time steps of 32,000 years (i.e., 5000 time steps), adopting the $\Gamma=-1.35$ IMF slope.  (This time range is much longer than $\sim$30-50~Myr ages of even the lowest mass RSGs.) The final distribution function was then simply summed under
the assumption of continuous star formation.

\subsection{Observed}
\label{Observed}

To construct the observed luminosity functions, we start with the recent M31 and M33 RSG candidate lists given by \citet{2021AJ....161...79M}, but modify what is included and the derived luminosities as described in this section.  Full details of the determinations of
luminosities from near-IR photometry are given in that paper, but we briefly summarize our basic procedure here.
The observed $J-K_s$ colors were first transformed to the standard \citet{BessellBrett} $J-K$ system.  Next the photometry was corrected for extinction. (Here we improve our treatment of extinction, as described below.)   The absolute magnitudes $M_K$ were then derived assuming true distance moduli of 24.40 for M31 (i.e., 760 kpc) , and 24.60 for M33 (830~kpc), following \citet{vandenbergh2000} and more recent references. The de-reddened colors were then used to derive temperatures and bolometric corrections to $M_K$ using the transformations derived from the MARCS stellar atmospheres \citep{Marcs75,Marcs92,marcs08}.

\subsubsection{Completeness}
\label{Sec-Sample}
For M31, we restrict our sample to RSGs with galactocentric distances $\rho$ less than 0.75. (The quantity $\rho$ is the de-projected distance within the plane of each galaxy, normalized to the radius at which the surface brightness $\mu_B=25.0$ mag arcsec$^2$; see  \citealt{RSGWRs} and \citealt{2021AJ....161...79M} for more details.)  Similarly for M33, we restrict our sample to $\rho<1.0$.  These limits were chosen for consistency with \citet{RSGWRs}, and correspond to the completeness limits of the optical photometry of the Local Group Galaxy Survey \citep{LGGSI} and the followup spectroscopic survey \citep{BigTable}.

In \citet{UKIRT} we constructed an M31 RSG luminosity function based upon our own Wide Field Camera (WFCAM) $J$ and $K_s$ images obtained on the 3.8 m United Kingdom Infrared Telescope (UKIRT) located on Maunakea, Hawai'i.  The survey covered 1.26 deg$^2$.  Subsequently, we used the Andromeda Optical and Infrared Disk Survey data ({\sc androids}) \citep{2014AJ....147..109S} to select RSGs over the entire 5 deg$^2$ of M31's optical disk \citep{2021AJ....161...79M}.  These data were taken at the Canada France Hawaii Telescope (CFHT), also located on Maunakea, with significantly superior seeing (0\farcs5-0\farcs6) to those of our UKIRT M31 data (1\farcs2-2\farcs6).  

When we compare the number of RSGs found in the regions of overlapping coverage, we find many more RSGs in the \citet{2021AJ....161...79M} CFHT data than in the \citet{UKIRT} UKIRT lists.  For instance, in the \citet{2021AJ....161...79M} CFHT data that corresponds to the UKIRT ``Field B" in \citet{UKIRT}, there are 1904 RSGs identified as more luminous than $\log L/L_\odot$=4.0; only 494 (26\%)  of these were found in the UKIRT survey.  

Careful examination revealed that the additional CFHT RSGs were present on the UKIRT images, but had been rejected as detections for several spurious reasons.   First, the UKIRT data for each field were obtained during each of two ``visits"; i.e., $J$ and $K_s$ images were both taken on one night, and then taken again on another night.  As described in \citet{UKIRT}, during each of the visits there were four dithered exposures for $J$ and for $K_s$.   In order for a detection to be considered valid in \citet{UKIRT}, it needed to be identified by the Cambridge Astronomy Survey Unit (CASU) pipeline software\footnote{http://casu.ast.cam.ac.uk/surveys-projects/wfcam} in both $J$ and $K_s$ in the same dither.  Furthermore, the star then had to be identified in both visits.  Due to the fact that the seeing was invariably worse in one visit than the other, this had the effect of losing some of the fainter stars.  Thus the conservative detection requirements, combined with variable (and poor) seeing,  accounted for about half of the losses. The explanation for the other half were issues with blending causing the CASU software to fail to notice the object as either a stellar or non-stellar source.  In contrast, \citet{2021AJ....161...79M} used standard IRAF {\sc daofind} for object detection and PSF-fitting {\sc daophot} for object detection and photometry for the CFHT detections.
These problems primarily affected the \citet{UKIRT} M31 luminosity function below $\log L/L_\odot$=4.5; above that, the normalized luminosity function is very similar to the one we derive here using the \citet{2021AJ....161...79M} data. 

We note that we do not expect these problems to affect the M33 luminosity function we derive from \citet{2021AJ....161...79M}, despite it also being based upon UKIRT WFCAM imaging.   Although those data also relied upon CASU pipeline,  they were taken as part of a search for AGB stars by \citet{2008A&A...487..131C} under much better seeing conditions (0\farcs7-1\farcs1) than our own M31 UKIRT data (1\farcs2-2\farcs6).  We also did not impose such draconian restrictions on matching; details are given in \citet{2021AJ....161...79M}.  Finally, M33 poses a far less crowded environment than does M31, as it is seen more face on, with an inclination of 56$^\circ$ \citep{1989AJ.....97...97Z} vs.\ 77$^\circ$ \citep{RC3}.

\subsubsection{Revisiting the Extinction Corrections}
\label{Sec-red}
In determining the extinction corrections for RSGs in M31 and M33,
\citet{2021AJ....161...79M} adopted a constant reddening for most stars ($A_V=0.75$ mag) but applied an
additional correction for the brightest RSGs ($K_s<14.5$)  i.e., $A_V=0.75+1.25\times(14.5-K_s)$  following \citet{UKIRT} and \citet{2021AJ....161...79M}.  This correction was based upon spectral fitting of 16 RSGs in M31 in \citet{MasseySilva}, and is illustrated in 
\citet{UKIRT}.   In retrospect, there were three problems with that approach.  First, the maximum $A_V$ of any of the stars included in the \citet{MasseySilva} was 2.5~mags, and, through
an unfortunate oversight, we extrapolated the relationship to brighter magnitudes, resulting in several unrealistically high extinction corrections. Second, the higher values were defined by only three stars, and the slope and break point ($K_s\sim14.5$) poorly determined. Third, there was no observational evidence that this same relationship should be applied to RSGs in M33.  We re-examine the issue
here, based upon (1) the additional spectral types now known for RSGs in M31 and M33, and
(2) our improved understanding of the average reddenings in these galaxies.

\begin{figure}[t]
\begin{centering}
\includegraphics[width=0.5\textwidth]{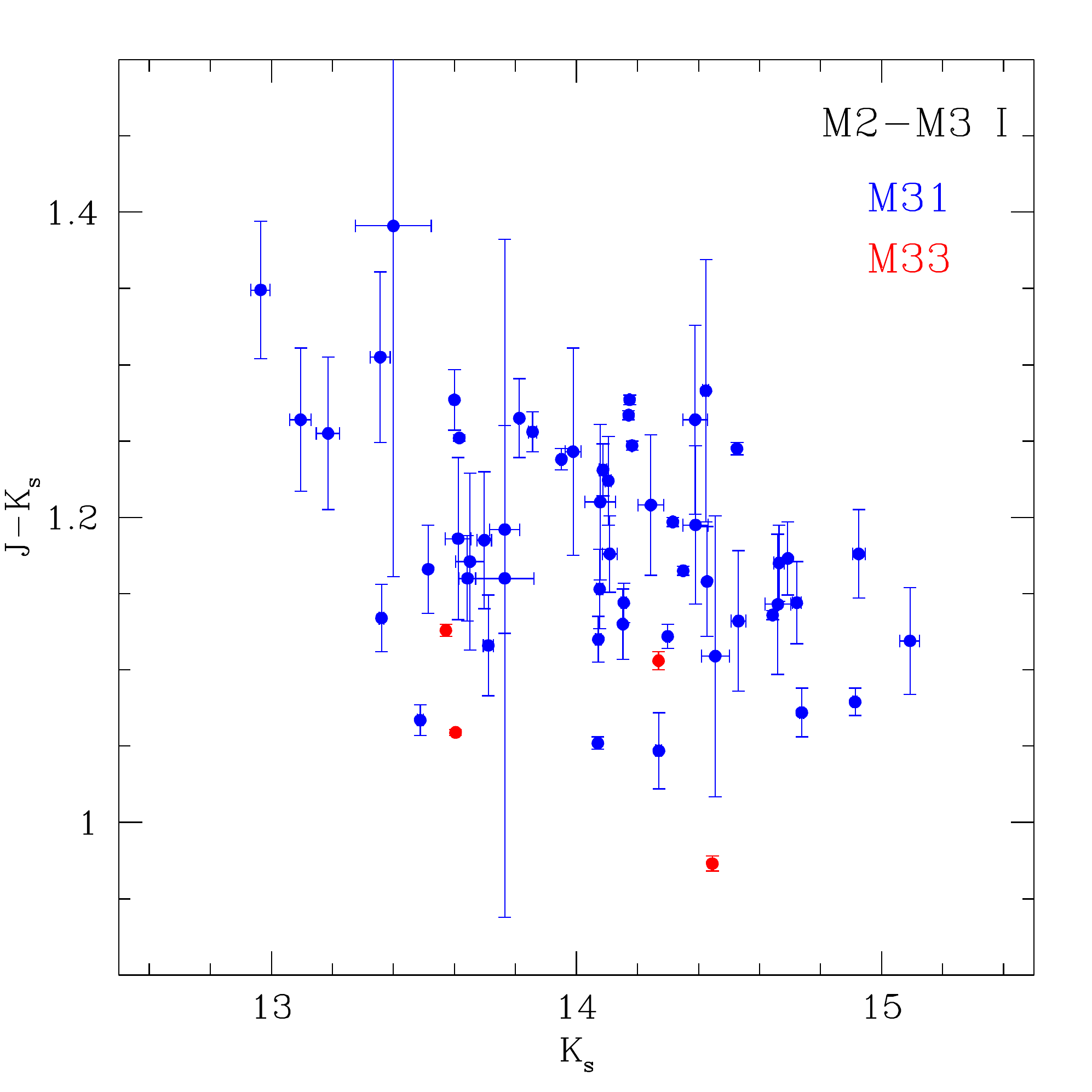}
\caption{\label{fig:M2s} NIR Colors of M2-3 Supergiants.  The $J-K_s$ colors are shown as a function of $K_s$ for known M2-3~I stars in M31 (blue) and M33 (red).  There is only a  weak correlation between the two for M31. In addition, the M31 M2-3 supergiants are systematically redder than their M33 counterparts, consistent with the average reddening being higher in M31.}
\end{centering}
\end{figure}

We begin our exploration of the reddening issue by examining the colors and previously derived physical properties of
a set of RSGs with similar spectral types. There are 54 M2-M3~I stars known in M31 and 4 in M33, which we list in Table~\ref{tab:M2s}.  Since these stars are of similar spectral type, we expect their
intrinsic $J-K_s$ colors to be the same\footnote{We even expect the colors to be similar for RSGs of the same spectral types between M31 and M33.  \citet{MasseySilva} showed that M2~I supergiants in M31 have very similar effective temperatures as RSGs of the same spectral types in the Milky Way, while \citet{EmilyMC} showed there was only about a 50~K difference in effective temperature between Milky Way RSGs and Magellanic Cloud RSGs of similar type.}. As shown in Figure~\ref{fig:M2s}, any variation with magnitude is
lost in the scatter. There is a weak linear correlation coefficient for the 54 M31 M2-M3 stars, with a value $r=-0.39$, but with a much smaller slope than suggested by the above relation:  From
the relationship above, we would have adopted $A_V=0.75$~mag at $K_s\sim15$, and $A_V=2.0$~mag 
at $K_s\sim13.5$, given that the color excess $E(J-K)$ will be $A_V/5.79$ \citep{Sch}.   Thus, we would have expected stars with $K_s\sim 13.5$ to be 0.2~mag redder than than stars with $K_s\sim15$, but that is not what we see.  (We will discuss the offset between the M31 and M33 stars further below, but the obvious interpretation is that the reddening is higher in M31.)  
Thus the extinction correction adopted by \citet{2021AJ....161...79M}, which was used in our preliminary luminosity function study \citep{UKIRT},  no longer appears to be justified. 
This is further emphasized by the fact that if the color correction based upon the above was applied, then the brighter M2-M3 stars would have temperatures of 4300-4450~K or higher,  rather than the 3600-3700~K one would expect given the spectral type \citep{EmilyMC,MasseySilva}.  
We are sure that there are RSGs in M31 with higher reddening, as found by \citet{MasseySilva} (due either to circumstellar material or located in a region of particularly high reddening), but {\it in general} the additional extinction correction for the brightest $K_s$ stars we previously used is unlikely to be correct.

Here we adopt the following approach.  For M31, we make use of the Panchromatic Hubble Andromeda Treasury (PHAT) survey \citep{PHAT}. Using this comprehensive photometry set, \citet{PHATDust} finds that the average extinction in the disk corresponds to an $A_V=1.0$~mag.  We can compare this to the extinction found by fitting MARCS RSG models to flux-calibrated spectrophotometry by \citet{MasseySilva}.  For the 16 M31 RSGs in their sample, the median $A_V$ value is 1.01~mag; the mean is 1.20~mag, with a 0.6~mag scatter, so this agreement is good. 

For M33, the extinctions for multiple regions was measured in a similar manner 
using the Panchromatic Hubble Andromeda Treasury: Triangulum Extended Region (PHATTER, \citealt{PHATTER}) data set  \citep{PHATTERDust}. The
average roughly corresponds to a value of $A_V=0.8$~mag (M. Lazzarini 2022, private communication). Although our recent studies have determined spectral types for many
RSGs in M33 (\citealt{2021AJ....161...79M} and references therein), the vast majority of these were obtained using multi-object fiber observations, where the accuracy of the flux calibration is poor.  However, we note that the $J-K_s$ colors of the M33 M2-3~I stars shown in Figure~\ref{fig:M2s} are less red than their M31 counterparts,  consistent with their having lower reddening.

We therefore adopt a constant reddening corresponding to $A_V=1.0$ for the M31 RSGs and $A_V=0.8$ mag for the M33 RSGs.  Since the extinction at $K$ is only 12\% of these
values (see, e.g., \citealt{Sch}), this corresponds to $A_K=0.12$ and $A_K=0.10$ for M31 and M33, respectively.  These corrections are applied after the minor transformation from $K_s$ to $K$ as given in \citet{Carpenter}, i.e., $K=K_s+0.044$.  Similarly we adopt color excesses of $E(J-K)=0.17$ and $E(J-K)=0.14$ before calculating the temperatures.  The equations we used for the color transformations of $J-K_s$ to $J-K$, conversion of colors to temperatures, and the calculation of bolometric luminosities can be found in Table 2 of \citet{2021AJ....161...79M} and references therein. 

We have no doubt that there are RSGs in M31 and M33 with significant additional extinction due to circumstellar material, as is known in the Milky Way and the Magellanic Clouds (see, e.g., \citealt{1969ApJ...158..619H,1986ApJ...302..675E,1999A&A...351..559V,1989AJ.....97.1120S,Smoke,2009AJ....137.4744L}), but we believe that adopting a constant reddening is an improvement over past attempts.  There is certainly a hint in Figure~\ref{fig:M2s} that the brightest $K_s$ stars are slightly redder.  This could be investigated further by fitting of well flux-calibrated optical data, and we have plans to do so.

\begin{figure}[t]
\begin{centering}
\includegraphics[width=0.6\textwidth]{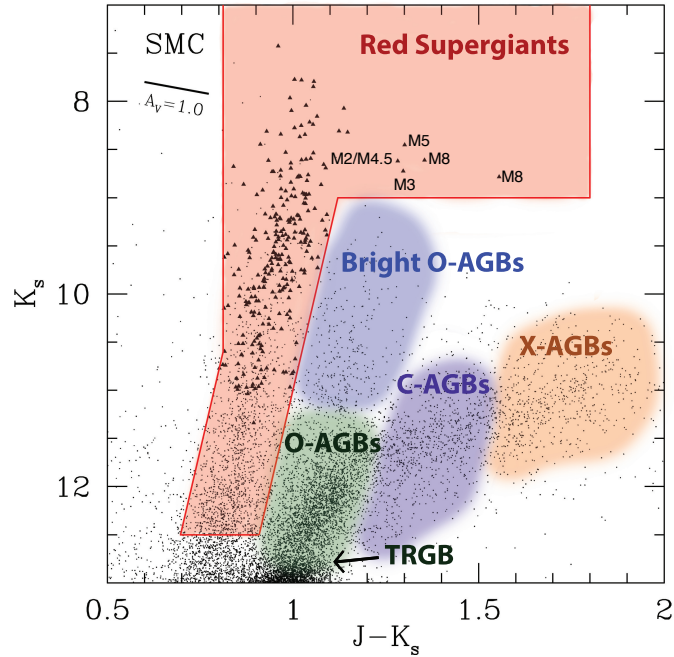}

\vskip -30pt
\caption{\label{fig:SMCWowSylvia} Color-magnitude diagram for the SMC RSGs and AGBs.  This figure shows the spectral types for the five SMC stars that are unexpectedly bright and red in the $(J-K_s,K_s)$ color-magnitude diagram.   The two M8 stars are classified as giants or bright giants by \citet{Dorda2018}, although their luminosities suggest otherwise.  The star labeled ``M5" is HV~11417, the peculiar M supergiant described in detail by \citet{1980ApJ...242L..13E}. It was suggested as a possible Thorne-Zytkow Object by \citet{2018MNRAS.479.3101B}, but described as likely Galactic halo object by \citet{2020ApJ...901..135O}. The
M3 I star is HV 2112, the TZO candidate identified by \citet{TZO} (see also \citealt{2018MNRAS.479.3101B} and \citealt{2020ApJ...901..135O}). The M2/M4.5 I star is
[M2002] SMC055188 \citep{Massey2002}, and identified as a spectrum variable by \citet{EmilyVariables}.  This figure is based upon the data given in Table 5 in \citet{RSGWRs} and is a modified version of their Figure 15.}
\end{centering}
\end{figure}

\subsubsection{The Reddest Stars}
\label{Sec-AGBs}
Another uncertainty we faced in \citet{UKIRT} and \citet{2021AJ....161...79M} was how to treat a group of stars that are in an unexpected location in the CMDs.  Invariably in our previous studies we have found stars in the ($J-K_s, K_s$) plane which are as red as AGB stars, but brighter than we expect AGBs to be found.  This is well illustrated in the color-magnitude diagram (CMD) of the Small Magellanic Cloud (SMC) shown in Figure~\ref{fig:SMCWowSylvia}.   In our studies of the RSG content of the LMC, M31, and M33 \citep{UKIRT,LMCBins,2021AJ....161...79M} we  {\it assumed} that these are heavily reddened RSGs, and determined extinction corrections for each star by projecting back along a reddening line to the median relation between $J-K_s$ and $K_s$ in our sample.  However,  by the time we studied RSG population of the SMC \citep{RSGWRs}, our spectroscopy was sufficiently complete to show that most of these stars really were as cool as their $J-K_s$ indicated, and were not the result of high reddening.  We illustrate this in Figure~\ref{fig:SMCWowSylvia} by indicating the spectral types
of these stars\footnote{Note that one of these stars is HV~2112, the well-known Thorne-\.{Z}ytkow object (T\.{Z}O) candidate \citep{TZO}.  \citet{2018MNRAS.479.3101B} has argued that the star is simply a super-AGB star, and suggested their own
T\.{Z}O candidate, HV~11417,  which is located in the same region, but see \citet{2020ApJ...901..135O}.}.

\begin{figure}
\begin{centering}
\includegraphics[width=0.42\textwidth]{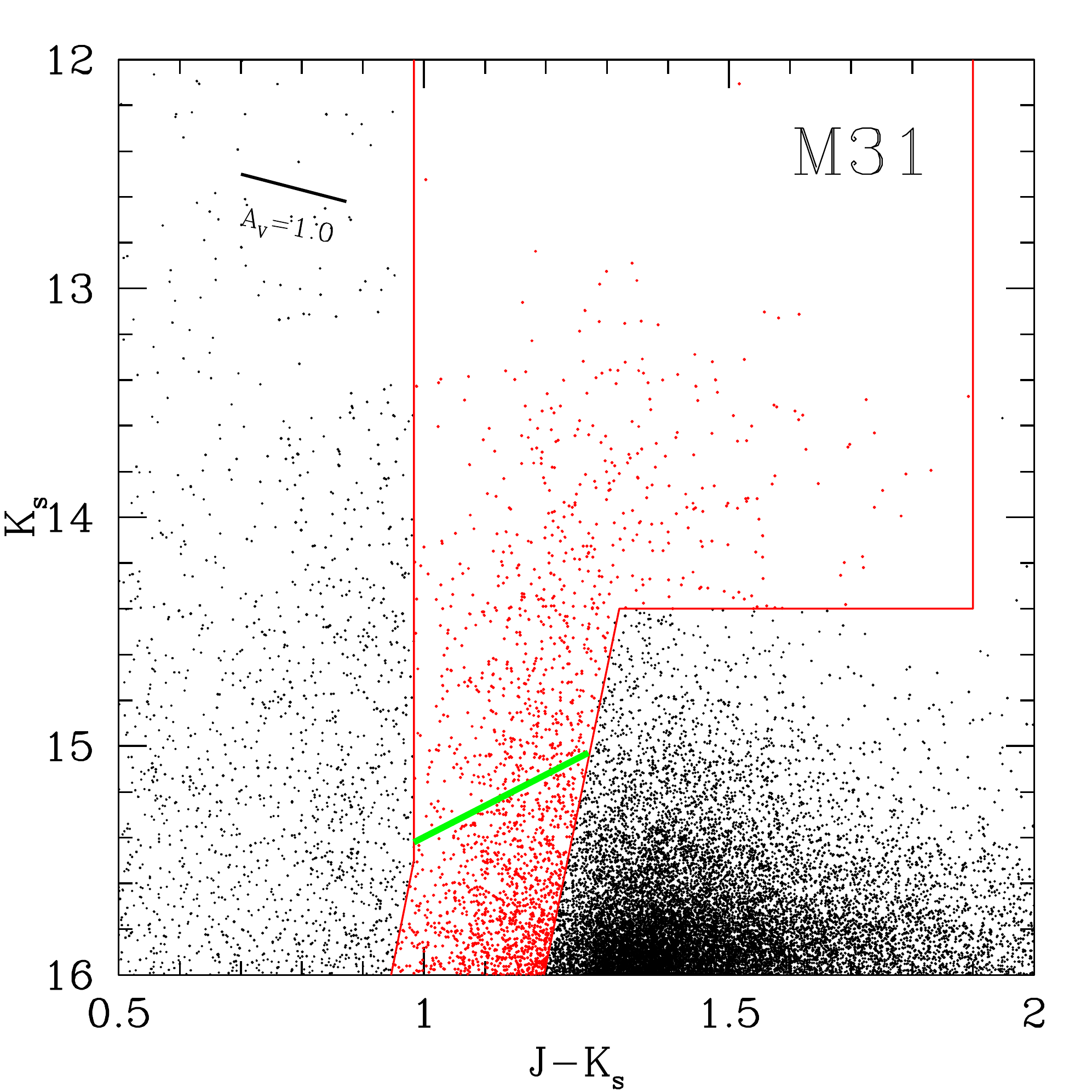}
\includegraphics[width=0.42\textwidth]{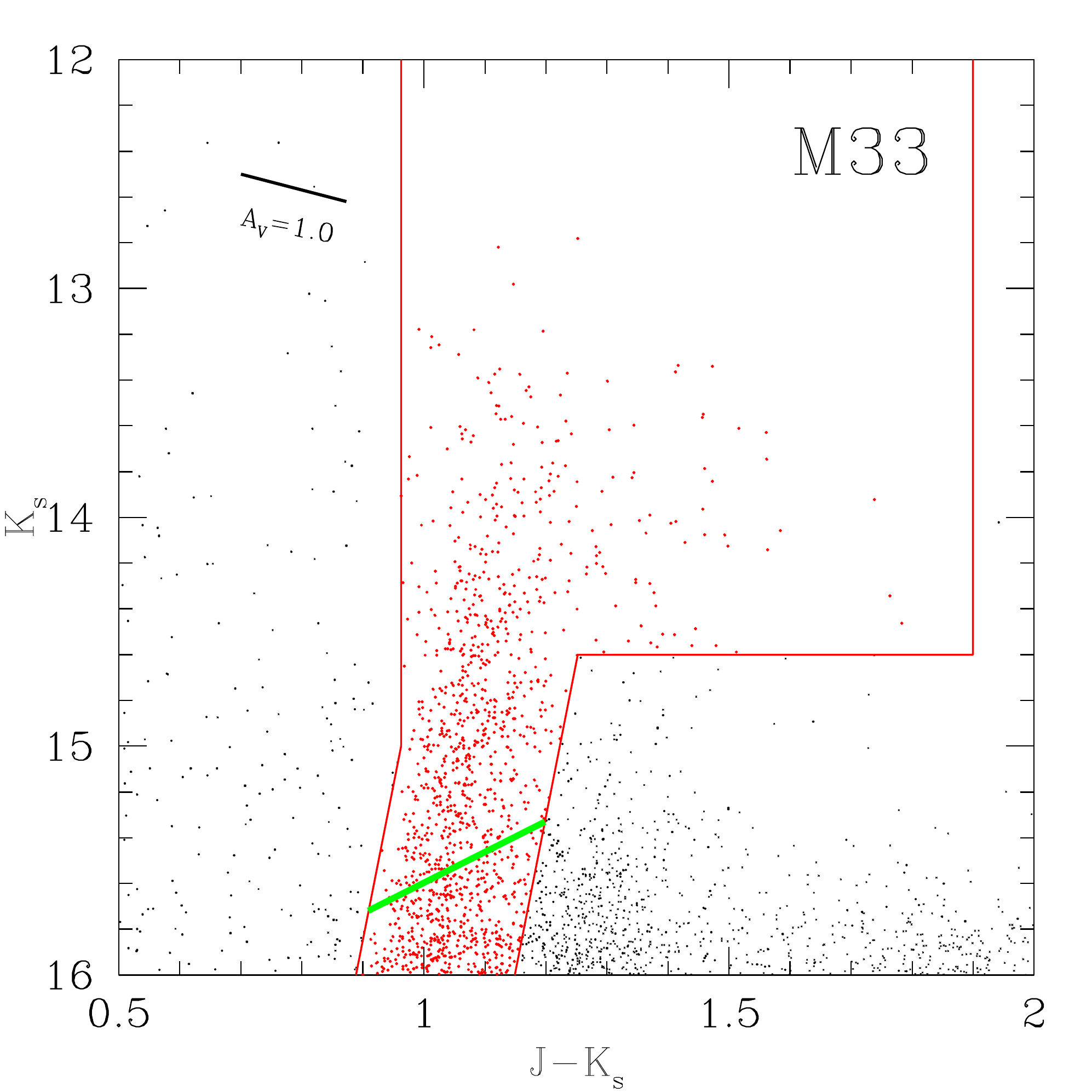}
\includegraphics[width=0.42\textwidth]{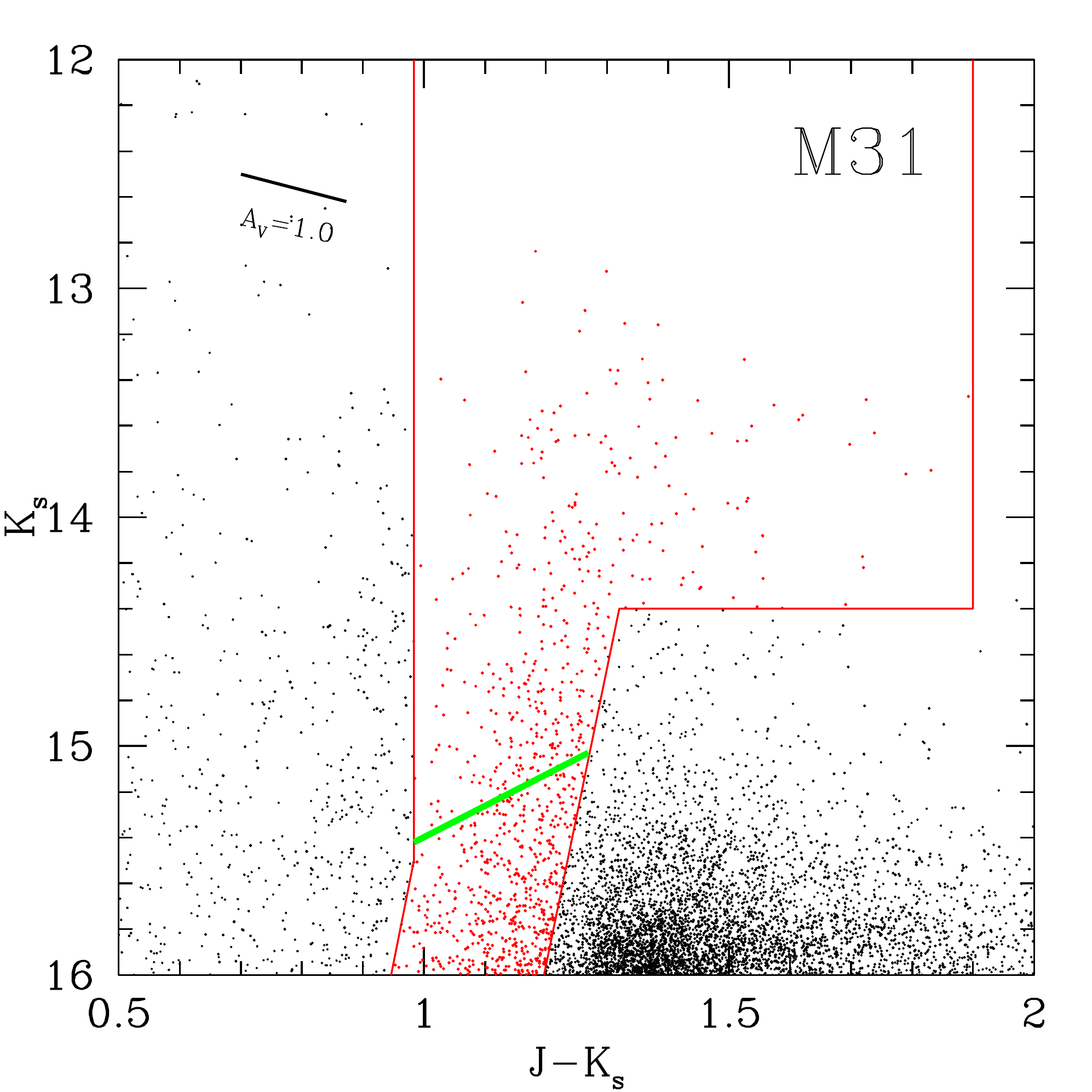}
\caption{\label{fig:M3133cmd} Color-magnitude diagrams for M31 and M33.  
The top two figures are
modified from those in \citet{2021AJ....161...79M}, with the additional restrictions of
$\rho\leq0.75$ for M31 and $\rho\leq1.00$ for M33.  Foreground stars were removed using
DR2 Gaia data in \citet{2021AJ....161...79M}; we have also removed the additional stars
found to be non-members using DR3 here.  The third panel, at bottom, shows the CMD for M31 restricting the sample to the regions with lower reddenings.
 The green diagonal line corresponds to 
$\log L/L_\odot=4.5$.  A reddening vector is shown at upper left in each panel.  In comparing with the CMD for the SMC shown in Figure~\ref{fig:SMCWowSylvia}, recall that there is $\sim$5.5~mag difference in distance modulus.
}
\end{centering}
\end{figure}

That said, the CMDs of M31 and M33 given in Figure~\ref{fig:M3133cmd} show some extremely red
stars ($J-K_s>1.5$).  If these stars have only typical reddenings, then their temperatures
are 3400~K, corresponding to a spectral subtype a little latter than an M5 supergiant \citep{Levesque2005}.  These are surprisingly cool for RSGs, and we suspect that these are primarily super-AGB stars (see, e.g., \citealt{2005A&A...438..273V} and \citealt{2017MNRAS.465..403G}).  We note that the number of stars eliminated is quite small: 52 stars out of 728 more luminous than $\log L/L_\odot$=4.5 for M31, and 8 out of 836 for M33.  One of these ``too red" M31 objects proved to be a known background galaxy.  We have excluded these 
stars with $J-K_s>1.5$ from our discussions of the luminosity function.  We note that by doing this we may be excluding some very dusty RSGs, such as the well-known
WOH G64 \citep{1986ApJ...302..675E,1998Ap&SS.255..405V,2009AJ....137.4744L} in the LMC. However, we expect these to be scant in number, and that most of these very red stars will prove
to be AGBs.

\subsubsection{Removing non-RSGs}
\label{Sec-Gaia}

There are three potential sources of contamination in the photometry lists of RSG candidates of \citet{2021AJ....161...79M}: foreground red stars (mostly dwarfs), background galaxies, and globular clusters.  Here we remove the residual contaminates before constructing our luminosity functions. 

As described in detail by \citet{2021AJ....161...79M}, for many years the primary way to separate extragalactic RSGs from foreground Galactic red dwarfs was via radial velocity studies, aided by ground-based parallaxes and proper motions for the brightest interlopers (see, e.g., \citealt{MasseyRSGs,MasseySilva,DroutM33,Massey2016}.)  This all changed with the availability of high precision Gaia astrometry, as was discussed extensively in \citet{ErinLBV}. (For an overview of Gaia, see \citealt{GenGaia}.) The RSG catalogs given in \citet{2021AJ....161...79M} were cleaned on the basis of parallax and proper motion data given in Gaia Data Release 2 \citep{DR2}.  Since then, improved values have become available to the community via Data Release 3 \citep{EDR3,DR3}.    We used these as follows:  First, we used the RSGs designated as
members by  \citet{2021AJ....161...79M} to determine the expected values for the parallaxes and proper motions, as there are zero-point issues that are dependent upon
location in the sky.  For M31, we find median values of  0.00 mas, +0.012 mas yr$^{-1}$, and $-0.077$ mas yr$^{-1}$ for the parallax and proper motions in right ascension and declination, respectively, using a sample of 1807 stars after eliminating outliers.  For M33, the median values are $-0.04$ mas, +0.059 mas yr$^{-1}$, and +0.020 mas yr$^{-1}$ using a sample of 1559 stars after eliminating outliers. We used these values to eliminate any previously assumed member if (a) their parallax was $>$$3.5\sigma$ than the median values, or (b) if either of their proper motion values were further from the expected (median) values by 3.5$\sigma$. For $\sigma$, we used the Gaia uncertainties in the relevant values unless they were smaller than the median uncertainty values (0.3~mas for the parallaxes and 0.3~mas yr$^{-1}$ for each of the proper motions).   If the stated uncertainties were smaller than the median uncertainties, we adopted the median uncertainty as a better representation.  This was done in order to avoid the situation of a star with values consistent with membership being rejected simply because the uncertainty was under-estimated. 

In addition to the improved astrometric data, we also reviewed the existing radial velocity data for these stars.   For M31, such radial velocity information provides a particularly powerful tool, as the galaxy's systemic velocity is $-300$ km s$^{-1}$ and has a rotational velocity of $\sim$240 km s$^{-1}$ (\citealt{MasseySilva} and references therein).  Thus, over most of the galaxy the separation is very clean, except for an ``alligator jaws" shaped area in the NE section, as shown in Figure 5 of \citet{DroutM31}, where the rotational velocity essentially cancels the systemic velocity.  For M33 the systemic velocity is $-180$ km s$^{-1}$ and the rotational velocity is 74 km s$^{-1}$ (\citealt{DroutM33} and references therein). 
To compare with the expected radial velocities as a function of position, we used the relationships between radial velocity and location found by \citet{DroutM31} and \citet{DroutM33} for M31 and M33, respectively.  The radial velocities for the M31 RSGs came for \citet{Massey2016}, while
the ones for M33 were determined by \citet{DroutM33}, supplemented by the project
described by \citet{2017AAS...22943304B}; the actual values of the latter are still in preparation.  In all cases the comparison agreed with the conclusions drawn from the Gaia data.  
The only star eliminated from our list based on published radial velocities (see below) was J004248.37+412506.5, which   \citet{DroutM31} designated a foreground object based on its radial velocity.  There are no Gaia measurements for the object, but we measured a radial velocity of -45.4 km s$^{-1}$, which is highly discrepant with the expected disk velocity at that location, $-216$~km s$^{-1}$.   The object is now considered a cluster (see, e.g., \citealt{2010ApJ...725..200F, 2013ApJ...769...10J, 2015ApJ...802..127J, 2016ApJ...824...42C}), known as Bol~129.    

Indeed, globular clusters at the distance of M31 and M33 are hard to distinguish from individual stars in many cases, as they are nearly stellar in angular size in typical ground-based seeing. We have therefore vetted our RSG list against catalogs of M31 and M33 clusters.  For M31 we used the on-line Version 5 of
the Revised Bologna Catalogue of M31 Globular Clusters and Candidates\footnote{http://www.bo.astro.it/M31/} \citep{2004A&A...416..917G}, supplemented by the 
on-line catalog\footnote{https://oirsa.cfa.harvard.edu/signature\_program/} of Hectospec M31 radial velocity of identified objects made available by N. Caldwell (2022, private communication).  The Hectospec data also allowed us to identify several additional foreground stars that had lacked Gaia data. 
As a bonus, both resources contained background galaxies that had been identified on the basis of their radial velocities.  \citet{2021AJ....161...79M} had eliminated many background galaxies by visual inspection of the images, but of course radial velocities provided a powerful additional means.  

Of the 38,856 M31 RSG candidates originally listed by \citet{2021AJ....161...79M}, the cross-identification with clusters eliminated 406 (1\%) spurious objects in M31.  The additional vetting using DR3 eliminated another 548 objects, an additional 1.4\%, mostly, however, among the brighter objects, as the faint
RSG candidates generally lack Gaia data.

There are fewer confirmed clusters in M33 than in M31.  We used the updated, on-line
version of the  \citet{2007AJ....134..447S} catalog available through Vizier\footnote{J/AJ/134/447/table3}.  For the clusters not already confirmed from high spatial resolution imaging with {\it HST}, \citet{2007AJ....134..447S} used good-seeing (0\farcs5) CFHT images to examine objects previously called clusters, and
classified them as either clusters, background galaxies, stars, or unknowns.  There were
29 matches with the list of 7088 RSG candidates of \citet{2021AJ....161...79M}, i.e., 0.4\%. This fraction is so much lower than that in M31 likely because most of the clusters in M33 are of intermediate age, and not true globular clusters (\citealt{vandenbergh2000} and references therein); hence they are not as red, and less likely to be confused with RSGs.  We included the ``unknowns" in culling process, as their number are scant.  Use of the DR3 eliminated another
251 (3.6\%).

We also addressed one additional deficiency with our previous methodology.  Since our study had excellent astrometry, we had previously cross-matched our lists with Gaia insisting on a match with 1\farcs0 using Vizier.  However, a careful reading reveals that this tool performs the matching using the original epochs, rather than correcting the coordinates to a common epoch using the measured proper motions.  This would result in high proper motion stars falling outside the 1\arcsec\ search radius, and thus appearing to have no Gaia data.   For Gaia, the epoch are 2016.0.  For the majority
of our M31 data, the epoch is 2007-2009.  For M33, the majority of the data have an epoch of 2005.  However, both the M31 and M33 lists
were supplemented by 2MASS sources, which have an epoch of 1997-2001.  Thus the potential contamination by high proper motion stars might
well be an issue primarily for the brighter stars.  We therefore rechecked the matches for all alledged ``non-Gaia" stars in the M31 and M33 catalog using a large (15\arcsec) search radius and looked for matches after taking the Gaia-measured proper motions into account. This resulted in eliminating 15 high proper motion stars in M31 and 10 in M33.

\subsubsection{Reddening and M31: Redux}
\label{Sec-LowExt}

As discussed above, we adopt an average extinction for the RSGs in M31 corresponding to $A_V=1.0$, and note that this is consistent with the results of spectral fitting by \citet{MasseySilva}.  However, there is another potential concern: RSGs that are so highly
hidden by extinction within the disk of M31 that we fail to count them.  This could, potentially, affect the luminosity function we adopt, providing a bias against the less luminous stars. 
\citet{2021AJ....161...79M} noted that the percentage of matches between their NIR-constructed catalog of RSGs and the optical Local Group Galaxy Survey (LGGS, \citealt{LGGSI,BigTable}) was much lower for M31 than for M33.  They showed that the fraction of matches were strongly correlated with the dust maps of \citet{2014ApJ...780..172D}.   Thus, in determining the binary frequency of RSGs in M31 and M33, \citet{M31M33RSGBins} restricted her M31 sample to those in regions of low extinction.
Similarly, in comparing the relative number of RSGs and Wolf-Rayet (WR) stars as a function of metallicity amongst Local Group galaxies, \citet{RSGWRs} used the same subset of RSGs drawn from the regions of low extinction.  We follow the same example here, with the RSGs drawn from regions in M31 where the matches with the LGGS are 50\% or greater. This reduces the number of RSGs
in our M31 sample from 804 to 409.  Applying the same criteria to the broader photometry
in \citet{2021AJ....161...79M} results in the third CMD shown Figure~\ref{fig:M3133cmd}. 

One bonus of this additional filtering is that potential contamination of the RSG sample by AGBs is reduced.  As shown in Figure 10 of \citet{2021AJ....161...79M}, the RSG population in M31 is mostly found in the well-known star formation ring, where most of the H\,{\sc i}, OB associations, and H$\alpha$ emission is found.  By contrast, the much older AGB population is more evenly distributed across the disk, but proportionately higher in the inner regions,
where the extinction is higher (see, e.g., Figure 9 in \citealt{2021AJ....161...79M} and Figure 2 in \citealt{2014ApJ...780..172D}).

\section{Results}
In the previous section, we described in detail how we computed the predicted RSG luminosity distributions
from the evolutionary models.  We also give details of the slight adjustments we made in the content and luminosities of the M31 and M33 RSGs identified by \citet{2021AJ....161...79M}.  Now we compare these two to determine 
if we can truly use the RSG luminosity function to constrain the time-averaged mass-loss rates of RSGs.  We finish by using our revised lists to consider the values for the highest luminosities of RSGs in M31 and M33,
and compare these values to those of RSGs in the Magellanic Clouds.

Since we have made multiple improvements in the M31 and M33 censuses as described in the previous sections, we are republishing these here as
Tables~\ref{tab:M31RSGs} and \ref{tab:M33RSGs}.    Following these improvements in the M31 and M33 RSG lists, we have reconsidered both membership and luminosities to our SMC and LMC RSG censuses (given in \citealt{RSGWRs} and \citealt{LMCBins}, respectively), and reissue our catalogs here as Tables~\ref{tab:SMCRSGs} and \ref{tab:LMCRSGs}, as we will use this information below\footnote{We adopted $A_V$=0.4~mag for the SMC RSGs and $A_V$=0.7 for the LMC RSGs based on the median reddenings found by \citet{EmilyMC}.  Checking for high proper motion stars that we incorrectly took as lacking Gaia data, we excluded
67 stars from the SMC list, and 10 from the LMC list, mostly at the bright end. We note that none of this affects the conclusions given by \citet{RSGWRs} concerning the relative number of RSGs and WRs.   For instance, of the 283 SMC RSGs listed in their Table 2 of having $\log L/L_\odot \ge 4.5$, a total of 10 stars (3.6\%) were listed incorrectly due to high proper motion.  This would change the ratio of the number of RSGs to WRs from 23.6 to 22.8, well within the quoted uncertainties.}.

\subsection{Comparison with the Evolutionary Models}
\label{Sec-doit}
\citet{UKIRT} used M31 to demonstrate that the RSG luminosity functions computed from the solar-metallicity \citet{Sylvia} evolutionary models were in reasonable agreement with the observations, while those computed from 10$\times$ and 25$\times$ higher RSG mass-loss rates could be ruled out, as they produced luminosity distributions that were much steeper than the observations (i.e., proportionally fewer high luminosity RSGs).  Here we address
a related question: are the observed luminosity functions also consistent with lower RSG mass-loss rates?

\begin{figure}[t]
\begin{centering}
\includegraphics[width=0.85\textwidth]{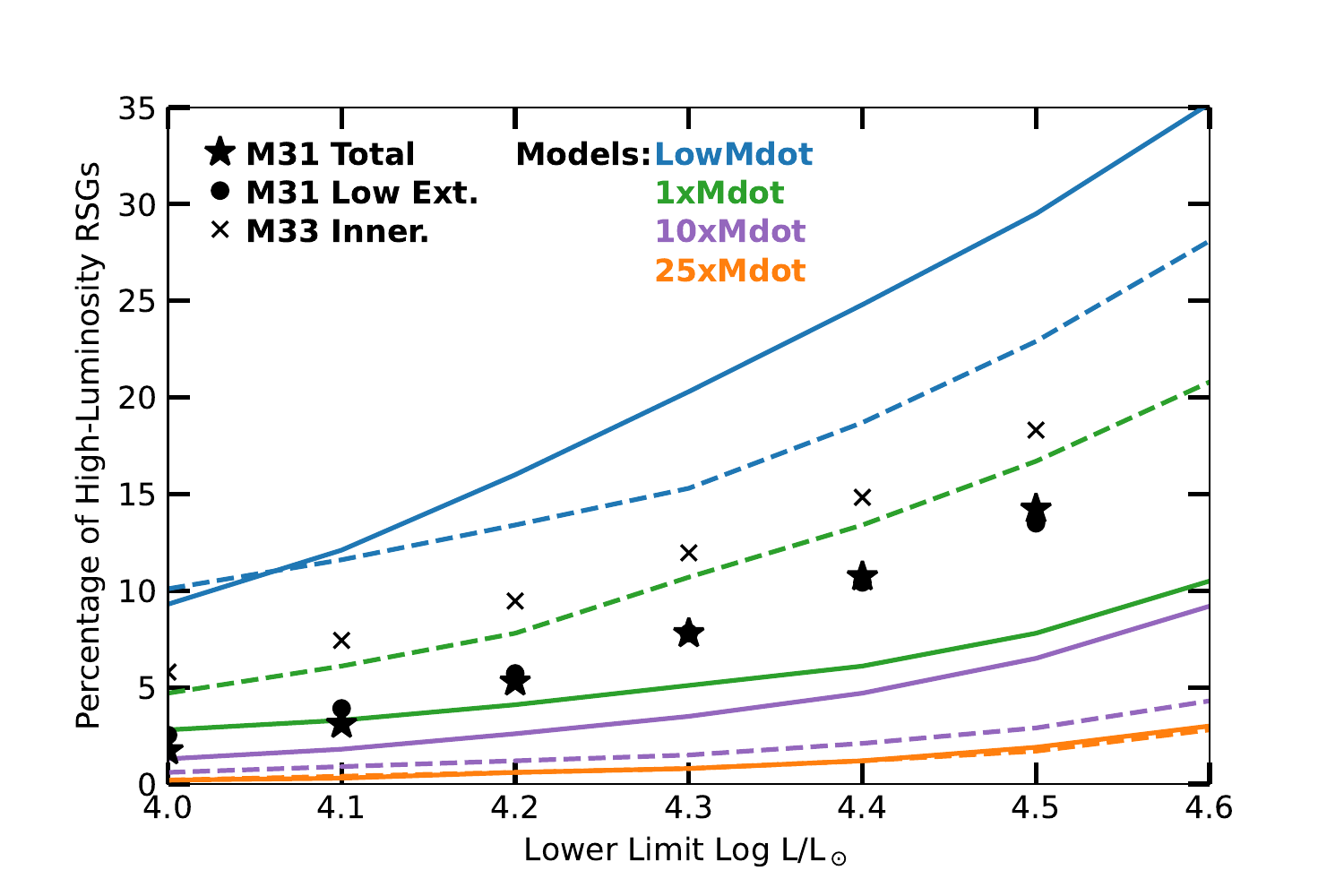}
\caption{\label{fig:percent} The percentage of  the highest luminosity RSGs ($\log L/L_\odot \geq 5.0$) observed compared to that predicted by the models with lower luminosity cutoffs from $\log L/L_\odot=4.0$ to 4.5.   The colored lines show the
predictions of the Geneva models; solid lines show the results from models that include rotation (initial value 40\% of the breakup speed) while the dashed
lines show the results from models without rotation. The points show the observed ratios.  The data are most consistent with the normal mass-loss rates (green) from \citet{Sylvia}, which includes the enhanced mass-loss during the supra-Eddington phase.   Higher mass loss-rates (purple and orange) predict too few high luminosity RSGs, while removing the enhanced mass-loss during the supra-Eddington phase (blue lines) predict too many high luminosity RSGs.
}
\end{centering}
\end{figure}

As discussed earlier, the low and normal mass-loss rate models were computed similarly, with the exception being that the low mass-loss 
models do not account for the supra-Eddington phases.  We expect that the primary effect of removing the enhanced mass-loss will be to increase the numbers of  the highest luminosity
RSGs, as high mass-loss rates will result in the most luminous RSGs shedding their hydrogen-rich outer layers and evolving back towards higher temperatures before undergoing core collapse.  Thus, the percentage of high luminosity RSGs compared to the entire sample should
decrease with increasing mass-loss rates.  To investigate this, we have computed the percentage of RSGs with high luminosities ($\log L/L_\odot \geq 5.0$) RSGs compared to the samples with $\log L/L_\odot$$\geq$4.0-4.5.  We show the theoretical predictions as colored lines in Figure~\ref{fig:percent} and give the calculated ratios in Table~\ref{tab:Fractions}.  We see from these that as the assumed RSG mass-loss rates increase,
the percentage of high luminosity RSGs indeed drops.

How do these model predictions compare to the observations? For comparison with the output of
{\sc syclist} we count only RSGs with effective temperatures $\leq$4200~K, but this restriction makes negligible difference in the fractions compared to, say, 4300 or 4400~K. We include the measured fractions as discrete points in Figure~\ref{fig:percent} and tabulate them in Table~\ref{tab:Fractions}.  For M31 we include both the full sample as well as the sample with low extinction.  We see that the results are very similar for each of these two samples.  {\it A comparison with the model predictions makes it clear that the fractions of high-luminosity RSGs are far more consistent with the mass-loss prescription used in the \citet{Sylvia} models which includes the enhanced mass-loss during the supra-Eddington phases.}  
For instance, we find that the percentage of high luminosity ($\log L/L_\odot \geq 5.0$)
RSGs in a sample with $\log L/L_\odot \geq 4.2$) is 5.7-7.8\%.  The low mass-loss models predict a value of 13.4-16.0\%, while the normal mass-loss models
predict ratios of 4.1-7.8\%.  Higher mass-loss models predict lower fractions.  The same trend is observed for all the luminosity cutoffs, although the M31 points are a little low compared to the models for the $\log L/L_\odot = 4.0$ and 4.1 points.  This suggests to us that the lowest luminosity points suffer some AGB contamination despite the efforts of \citet{2021AJ....161...79M} to separate the RSG and AGB populations in the CMDs.

We also include in Table~\ref{tab:Fractions} the results for the inner portion ($\rho <0.25$) of M33, where the metallicity is expected to be solar or slightly lower.
There we find slightly higher fractions than the normal mass-loss rates predict, but still in far better agreement with those than with lower mass-loss rates.
We attribute this small discrepancy to the lower metallicity of the M33 sample compared to that of M31, as lower metallicities will favor longer life-times for the higher luminosities RSGs, which might evolve to WRs were the metallicities higher.

We conclude that the enhanced RSG mass-loss that is included during the surpa-Eddington phases is not only justified from a theoretical point of view, but results in predictions that are a close match to the observations based on mass-limited samples in M31 and M33.  Lower mass-loss rates predict too many high-luminosity RSGs compared to observations.

\subsection{The Most Luminous RSGs}
\label{Sec-HDL}

Besides achieving our primary purpose, we can also use our knowledge of the M31 and M33 luminosity functions to answer another question of astrophysical importance: how does the luminosity limit of RSGs change with metallicity?  

In a landmark paper, \citet{1979ApJ...232..409H} showed although there were hot, massive stars with luminosities as high as $\log L/L_\odot\sim6.3-6.7$ ($M_{\rm bol}=-11$ to $-12$),  there were no cooler massive stars with luminosities above $\log L/L_\odot = 5.7-5.9$ ($M_{\rm bol}=-9.5$ to $-10$), roughly corresponding to 40$M_\odot$.  This ``Humphreys-Davidson limit" decreased with decreasing temperatures for the hottest stars, and then flattened to a constant value for stars cooler than about 15,000~K.  This was surprising, as the evolution of massive stars is expected to take place at nearly constant luminosities: what was stopping the most luminous hot stars to evolve to cooler temperatures?

\begin{figure}[t]
\begin{centering}
\includegraphics[width=0.9\textwidth]{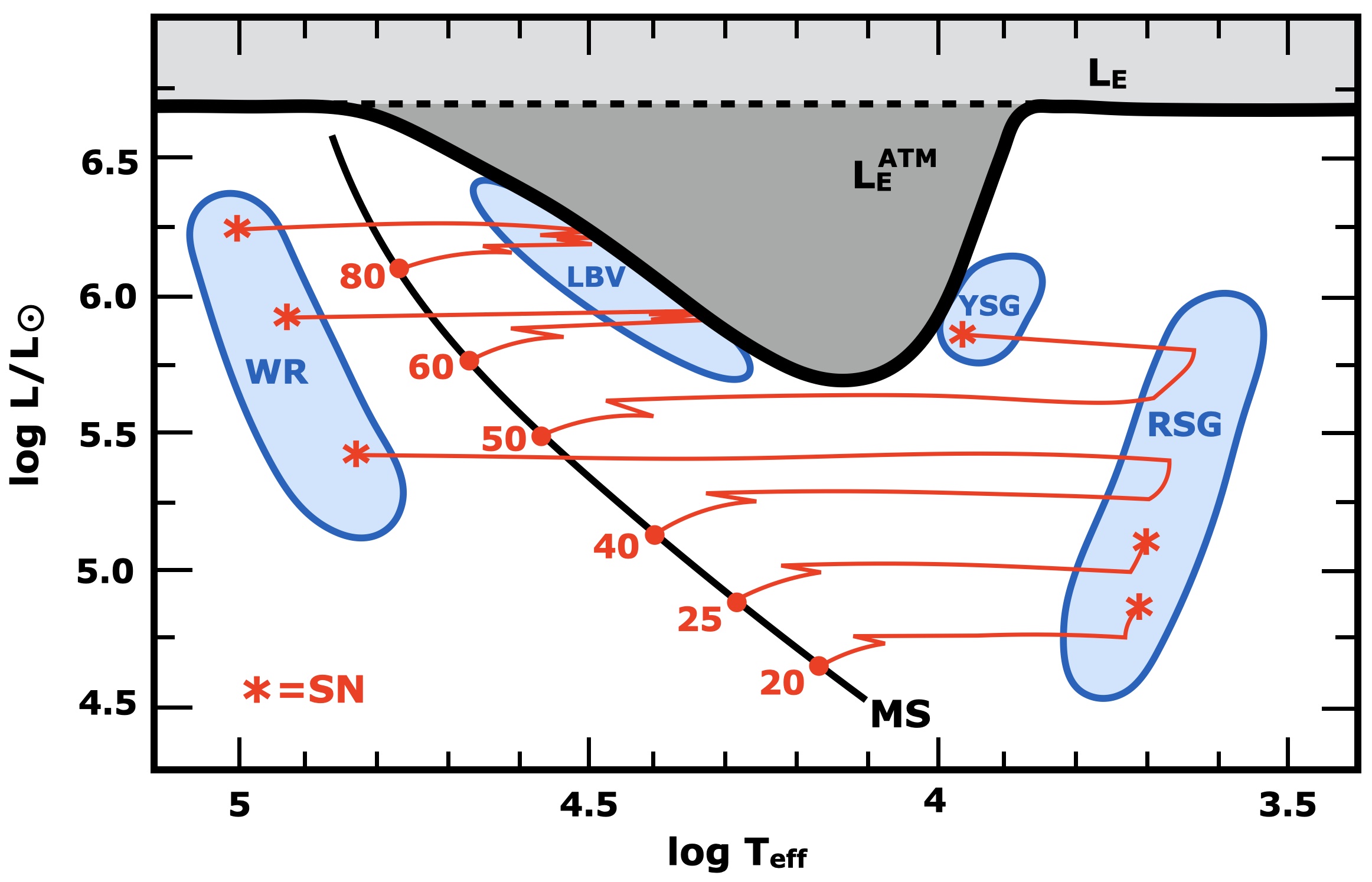}
\caption{\label{fig:Emily}  Schematic diagram showing the location of the Eddington trough.
 Stars evolve from the main-sequence (MS)
Higher luminosity (mass) stars encounter the ``atmospheric Eddington Limit" (L$^{\rm ATM}_E$) and enter a Luminous Blue
Variable (LBV) phase before evolving to Wolf-Rayets (WR).  Stars of lower luminosity (mass) evolve to the RSG phase.  The
most luminous of these will turn around and evolve back to yellow supergiants (YSGs). This figure is based on Figure 22.2 
from \citet{LamersLevesque}, and kindly provided to us by Emily Levesque. 
}
\end{centering}
\end{figure}

The upper luminosity of stars, known as the Eddington limit, is set primarily by the outward force generated by radiation pressure and the inward force of gravity. Classically, the Eddington limit is calculated assuming that electron scattering is the only opacity source \citep{1926ics..book.....E}.  While this is a good approximation for hot, main-sequence stars, line opacity, particularly in the UV, becomes important as a massive star evolves and cools (see, e.g., \citealt{1988ApJ...324..279L}).  The result is that the upper luminosity limit should be a function of effective temperature, with a minimum value (the so-called ``Eddington trough") reached at an effective temperature of 15,000~K.  This is illustrated in Figure~\ref{fig:Emily}. Since stellar evolution proceeds at roughly constant luminosity, the upper luminosity of stars cooler than this, such as the yellow and red supergiants, is set by the lowest luminosity of this trough, just as \citet{1979ApJ...232..409H} found observationally. As stars reach this luminosity limit, high mass-loss rates causes them to lose their extended, convective outer layers, and the stars will shrink and heat up, moving to the left in the H-R diagram, as also shown in Figure~\ref{fig:Emily}.

Theoretically, what do we expect to happen to the Humphreys-Davidson limit and the Eddington trough as a function of metallicity? 
\citet{1998ApJ...504..200U} used LTE plane-parallel models to compute the expected
dependence of the Eddington trough as a function of metallicity.  Line opacity should be increasingly important at higher metallicities, and indeed their
calculations show that
at M31-like metallicities (1.5$\times$ solar, \citealt{Sanders}) this limit should be at about
$\log L/L_\odot\sim5.6$~dex, while at SMC-like metallicities (0.2$\times$solar, \citealt{1998RMxAC...7..202K,2009A&A...496..841H,2017MNRAS.467.3759T}) the limit should be $\sim$6.0~dex.   However, further studies have shown that the situation is more complicated than that, and that there is a competing effect that compensates for the metallicity-dependent mass-loss \citep{2020A&A...635A.175H,2021MNRAS.506.4473S}.  Excess mixing in the superadiabatic outer layers of a star will help keep the star compact, and force evolution back bluewards, limiting the upper luminosity of cooler supergiants.  This process is strongest at low metallicities, leading to the expectation that the upper luminosity limit should be around $\log L/L_\odot \sim$ 5.5 regardless of luminosity,
at least between SMC-like and solar metallicities.

Here we can investigate this afresh using our revised luminosities.  In Table~\ref{tab:M31M33MostLums} we list the most luminous RSGs
in M31 and M33, where the latter are divided by their galactrocentric distance within the plane of M33 and have metallicities assigned as done
by \citet{NeugentM33} and references therein.  We see that the upper luminosity of RSGs is $\log L/L_\odot\sim5.4$ regardless of metallicities.  This result
is in excellent agreement with that expected by \citet{2020A&A...635A.175H} and \citet{2021MNRAS.506.4473S}, and with the findings by \citet{2018MNRAS.478.3138D} and \citet{2020MNRAS.493..468D} that the RSG luminosity limit in the Magellanic Clouds is close to $\log L/L_\odot\sim5.5$. A recent study of the RSG upper luminosity
limit in M31 using Spitzer mid-IR photometry by \citet{2022MNRAS.510.3132M} found a similar limit.

The \citet{2018MNRAS.478.3138D} and \citet{2020MNRAS.493..468D} studies relied upon somewhat older catalogs of Magellanic Cloud RSGs; there have been significant improvements in our knowledge of the RSG content of both galaxies 
thanks to recent work (e.g., \citealt{2020A&A...639A.116Y, LMCBins, RSGWRs}).  In Table~\ref{tab:MCsMostLums} we list the most luminous Magellanic Cloud RSGs, where we have improved the treatment of luminosities from \citet{LMCBins} and \citet{RSGWRs} as stated earlier. In agreement with the \citet{2018MNRAS.478.3138D} study, we find that $\log L/L_\odot=5.5$ is a good approximation to the RSG upper luminosity limit, and have extended such studies to the higher
metallicity of M31, with results that agree with \citet{2022MNRAS.510.3132M} using a sample of RSGs identified independently of the data we use here.

We note that in all of these galaxies, the values of the luminosities could be improved somewhat by modern spectrophotometry and model
fitting.  However, while this work might shift the results by 0.1~dex or so, it is unlikely to discover any significant metallicity dependence to the upper luminosity limits of RSGs.  We also note that these lists of the most luminous RSGs are unlikely to be complete.  Dusty RSGs, such as WOH G64 \citep{1986ApJ...302..675E} and the recently discovered ``LMC3" \citep{2022arXiv220911239D} are missing from these lists, as their $J-K$ colors are so large. Nevertheless, the luminosities of these
stars are comparable with the upper limits found here: \citet{2009AJ....137.4744L} finds a value of
$\log L/L_\odot = 5.45\pm0.05$
for WOH G64, while
\citet{2022arXiv220911239D} finds
$\log L/L_\odot =5.47\pm0.07$.

\section{Summary and Conclusions}

Mass-loss in RSGs is generally recognized to be episodic (see, e.g., \citealt{1999A&A...351..559V}), but most RSG mass-loss prescriptions used in evolutionary calculations are based upon ``instantaneous" measurements that do not take these events into account. \citet{Sylvia} addressed this problem by increasing the prescribed mass-loss rates by a factor of 3 whenever the model's luminosity exceeded the Eddington luminosity by a factor of 5.  The effect of this was to increase the life-time averaged mass-loss of 20$M_\odot$ by
a factor of 10; there is little effect expected on RSGs of lower luminosities.  This potentially solves the ``red supergiant problem"
\citep{SmarttIIP}, as the increased mass-loss rates would result in higher mass (luminosity) RSGs evolving to higher temperatures before undergoing core collapse.    

In this paper we have checked the approach used by \citet{Sylvia} observationally. The effect of including the enhanced mass-loss rates during the supra-Eddington phases will be to decrease the proportion of the highest luminosity RSGs.  For comparison, we have used newly computed Geneva models that differ from the \citet{Sylvia} models only in not applying a supra-Eddington correction.  As expected, they predict a larger fraction of high-luminosity RSGs than the standard models which includes enhanced mass-loss during the supra-Eddington phases.  In order to compare the predictions of these models against observations, we have started with the RSG samples in M31 and M33 given by \citet{2021AJ....161...79M}, and made improvements to both the memberships and luminosities calculations.  We have shown that the extinction corrections made by \citet{2021AJ....161...79M} were a little too high based upon the inferred temperatures for M2-3~I RSGs. Instead, we have recomputed the luminosities by adopting a uniform reddening for each galaxy based on the average values from the M31 PHAT \citep{PHAT,PHATDust} and M33 PHATTER \citep{PHATTER,PHATTERDust} studies.  Given that we are using these $A_V$ values to compute corrections to $K$, the differences in luminosities are slight in comparison with \citet{2021AJ....161...79M}, except for the brightest RSGs where we have argued that \citet{2021AJ....161...79M}
over-corrected the data.  We further cleaned the sample using improved Gaia Data Release 3 results, and also removed known clusters.

We find that the percentage of high-luminosity RSGs in M31 is a good match to the predictions of the \citet{Sylvia} models.  In contrast, the low mass-loss models over-predict the percentage of high-luminosity RSGs.  Even lower rates, such as those proposed by \citet{2020MNRAS.492.5994B}, would require an even higher proportion of high-luminosity RSGs, contrary to what we find.  For comparison, we also have shown that the predictions from the Geneva models with 10$\times$ and 25$\times$ enhanced RSG mass-loss rates predict a much lower proportion of high-luminosity RSGs than what is observe. Thus, we find that these much higher rates can also be ruled out, in accordance with the conclusions of \citet{UKIRT}.

For this study we have relied exclusively on the Geneva models, which do not include the effect of binarity.  Yet we know that the binary fraction of RSGs is
$\sim$30-40\% in M31 and the inner portions of M33 \citep{M31M33RSGBins}.  In our earlier study, \citet{UKIRT} showed that there was little difference in the general shape of the RSG luminosity functions derived 
using the Geneva models and the binary BPASS models of \citet{2018MNRAS.479...75S}; the major difference was at the
highest luminosities, which we attributed to the BPASS models using the older, low-mass prescriptions.
The influence of binarity on RSG luminosities has been further discussed
by \citet{2021A&A...645A...6Z} in the context of the red supergiant problem.  They conclude that the expected number of high luminosity RSGs is independent of the inclusion of binary evolution.  There are, of course, other questions that we hope to pursue in future work.  For instance, how sensitive are the theoretical RSG luminosities to the prescriptions of main-sequence mass-loss?

Finally, we have used our knowledge of these RSG populations to identify the most luminous RSGs in M31 and M33, and we have compared these values to those for RSGs in the SMC and LMC.   Our data show that the most luminous RSGs have $\log L/L_\odot\sim5.4$ regardless of metallicity, in accord with the findings of \citet{2018MNRAS.478.3138D}.  It has long been supposed that the Humprheys-Davidson limit \citep{1979ApJ...232..409H} should be metallicity dependent \citep{1998ApJ...504..200U}, but more recent work \citep{2020A&A...635A.175H,2021MNRAS.506.4473S} has noted a competing effect: at lowest metallicities excess mixing will reduce the luminosities of the coolest RSGs since these stars will
be driven back to warmer temperatures.  Our knowledge of the RSG content of these galaxies supports this conclusion.

\begin{acknowledgements}
Lowell Observatory sits at the base of mountains sacred to tribes throughout the region. We honor their past, present, and future generations, who have lived here for millennia and will forever call this place home.  The data from \citet{2021AJ....161...79M} used in this study were all obtained on Maunakea, and we wish to recognize and acknowledge the very significant cultural role and reverence that the summit of Maunakea has always had within the indigenous Hawaiian community.  We are most fortunate to have used observations obtained from this mountain.

As this work progressed, we benefited from useful conversations and feedback during the bi-weekly Zoom meetings with Dr.\ Emily Levesque and her massive stars research group, particularly from Drs.\ Trevor Dorn-Wallenstein, Azalee Bostroem, and Dr.\ Levesque herself. We are also indebted to Dr.\ Levesque for providing us with Figure~\ref{fig:Emily}, an
updated version from the one that appears in the recent text on stellar evolution, \citet{LamersLevesque}, and to Dr.\ Dorn-Wallenstein for comments on an earlier version of the manuscript. 
We are also grateful to Dr.\ Nelson Caldwell for help in eliminating some of the clusters and background galaxies that were in the original sample of M31 RSG candidates, and to Dr.\ Margaret Lazzarini for calculating the average extinction of star-forming regions in M33 from \citet{PHATTERDust} for us.  We also thank  Dr.\ Jacco van Loon for a kind and useful 
referee report that resulted in multiple improvements in the manuscript.

Partial support for this work was provided by the National Science Foundation through AST-83116 awarded to P.M.  In addition, support for K.F.N. was provided from NASA through the NASA Hubble Fellowship grant HST-HF2-51516 awarded by the Space Telescope Science Institute, which is operated by the Association of Universities for Research in Astronomy, Inc., for NASA, under contract NAS5-26555. S.E., C.G., and G.M. have received funding from the European Research Council (ERC) under the European Union's Horizon 2020 research and innovation program (grant agreement No.\ 833925, project STAREX).

This work has made use of data from the European Space Agency (ESA) mission
{\it Gaia} (\url{https://www.cosmos.esa.int/gaia}), processed by the Gaia
Data Processing and Analysis Consortium (DPAC,
\url{https://www.cosmos.esa.int/web/gaia/dpac/consortium}). Funding for the DPAC
has been provided by national institutions, in particular the institutions
participating in the {\it Gaia} Multilateral Agreement.  

\end{acknowledgements}

\facilities{UKIRT (WFCam NIR wide-field camera), CFHT (WIRCam NIR wide-field camera), CTIO:2MASS}

\bibliographystyle{aasjournal}
\bibliography{masterbib.bib}

\begin{deluxetable}{l l c c c c}
\tabletypesize{\tiny}
\tablecaption{\label{tab:M2s} M2-3 Supergiants in M31 and M33}
\tablewidth{0pt}
\tablehead{
\colhead{Star\tablenotemark{a}}
&\colhead{Sp.\ Type}
&\colhead{$K_s$}
&\colhead{$\sigma_{K_s}$}
&\colhead{$J-K_s$}
&\colhead{$\sigma_{J-K_s}$}
}
\startdata
\hline
\multicolumn{6}{c}{M31}\\
\hline
J004312.43+413747.1 & M2 I      & 12.965 &0.031 &1.349 &0.045 \\
J004047.22+404445.5 & M2 I      & 13.096 &0.035 &1.264 &0.047 \\
J004035.08+404522.3 & M2 I      & 13.186 &0.038 &1.255 &0.050 \\
J004125.72+411212.7 & M2 I      & 13.357 &0.033 &1.305 &0.056 \\
J004148.74+410843.0 & M2 I      & 13.361 &0.008 &1.134 &0.022 \\
J004034.74+404459.6 & M2.5 I    & 13.400 &0.124 &1.391 &0.230 \\
J004451.76+420006.0 & M2 I      & 13.488 &0.010 &1.067 &0.010 \\
J004047.82+410936.4 & M3 I      & 13.514 &0.001 &1.166 &0.029 \\
J004447.74+413050.0 & M3 I      & 13.600 &0.001 &1.277 &0.020 \\
J004030.64+404246.2 & M3 I      & 13.612 &0.042 &1.186 &0.053 \\
J004252.78+405627.5 & M2 I      & 13.616 &0.001 &1.252 &0.002 \\
J004120.25+403838.1 & M1-2 I    & 13.643 &0.027 &1.160 &0.028 \\
J004219.25+405116.4 & M2 I      & 13.651 &0.047 &1.171 &0.058 \\
J004501.30+413922.5 & M3+ I     & 13.698 &0.024 &1.185 &0.045 \\
J004415.17+415640.6 & M2 I      & 13.711 &0.016 &1.116 &0.033 \\
J004124.81+411206.1 & M2 I      & 13.765 &0.096 &1.160 &0.222 \\
J004424.94+412322.3 & M3 I      & 13.765 &0.050 &1.192 &0.068 \\
J004118.29+404940.3 & M2 I      & 13.813 &0.004 &1.265 &0.026 \\
J004108.42+410655.3 & M3 I      & 13.856 &0.013 &1.256 &0.013 \\
J004030.92+404329.3 & M2 I      & 13.951 &0.006 &1.238 &0.007 \\
J004447.08+412801.7 & M2.5 I    & 13.989 &0.026 &1.243 &0.068 \\
J004244.52+410345.1 & M2 I      & 14.070 &0.003 &1.052 &0.004 \\
J004517.25+413948.2 & M2 I      & 14.072 &0.009 &1.120 &0.015 \\
J004604.18+415135.4 & M2 I      & 14.076 &0.011 &1.153 &0.026 \\
J004147.27+411537.8 & M2 I      & 14.078 &0.050 &1.210 &0.051 \\
J004120.96+404125.3 & M3 I      & 14.087 &0.013 &1.231 &0.017 \\
J004614.57+421117.4 & M3 I      & 14.104 &0.010 &1.224 &0.029 \\
J004606.25+415018.9 & M2 I      & 14.109 &0.024 &1.176 &0.025 \\
J004552.56+414913.4 & M2 I      & 14.152 &0.006 &1.130 &0.023 \\
J004026.79+404346.4 & M2 I      & 14.155 &0.009 &1.144 &0.013 \\
J004101.02+403506.1 & M3 I      & 14.171 &0.002 &1.267 &0.003 \\
J004558.92+414642.1 & M2 I      & 14.174 &0.002 &1.277 &0.003 \\
J004015.18+405947.7 & M3 I      & 14.182 &0.002 &1.247 &0.003 \\
J004144.65+405446.8 & M2 I      & 14.244 &0.042 &1.208 &0.046 \\
J004449.89+415855.8 & M2 I      & 14.270 &0.010 &1.047 &0.025 \\
J004507.87+413109.0 & M3 I      & 14.299 &0.006 &1.122 &0.008 \\
J004509.73+415950.1 & M2 I      & 14.316 &0.002 &1.197 &0.003 \\
J004211.63+405017.9 & M2 I      & 14.350 &0.002 &1.165 &0.003 \\
J004354.50+411108.3 & M3 I      & 14.389 &0.040 &1.264 &0.062 \\
J004112.38+410918.5 & M2 I      & 14.390 &0.042 &1.195 &0.052 \\
J004122.48+411312.9 & M2 I      & 14.424 &0.010 &1.283 &0.086 \\
J004531.31+415918.3 & M2 I      & 14.428 &0.002 &1.158 &0.036 \\
J004103.55+410750.0 & M2 I      & 14.455 &0.046 &1.109 &0.092 \\
J004213.75+412524.7 & M3 I      & 14.526 &0.003 &1.245 &0.004 \\
J004502.62+414408.3 & M2 I      & 14.531 &0.024 &1.132 &0.046 \\
J004019.15+404150.8 & M2 I      & 14.643 &0.002 &1.136 &0.003 \\
J004133.42+403721.1 & M3 I      & 14.660 &0.042 &1.143 &0.046 \\
J004412.67+413947.8 & M2 I      & 14.664 &0.016 &1.170 &0.025 \\
J004035.16+404105.2 & M2 I      & 14.693 &0.003 &1.173 &0.024 \\
J004623.75+420141.4 & M2 I      & 14.722 &0.015 &1.144 &0.027 \\
J004356.22+413717.6 & M2 I      & 14.738 &0.007 &1.072 &0.016 \\
J004032.89+410155.1 & M3 I      & 14.913 &0.007 &1.079 &0.009 \\
J004359.94+411330.9 & M2 I      & 14.925 &0.021 &1.176 &0.029 \\
J004328.69+410816.4 & M2 I      & 15.093 &0.032 &1.119 &0.035 \\ \hline
\multicolumn{6}{c}{M33}\\
\hline
J013319.13+303642.5 & M3 I       & 13.572 &0.001 &1.126 &0.004 \\
J013258.18+303606.3 & M2.5 I     & 13.604 &0.001 &1.059 &0.002 \\
J013305.17+303119.8 & M2-2.5 I   & 14.269 &0.003 &1.106 &0.006 \\
J013349.83+303224.6 & M2.5 I     & 14.445 &0.003 &0.973 &0.005 \\
\enddata
\tablenotetext{a}{Designation from LGGS \citep{LGGSI}.}
\tablecomments{Spectral types and photometry are from \citealt{2021AJ....161...79M} and references therein.}
\end{deluxetable}

\rotate
\begin{deluxetable}{l l l l l l l l l l l l l l l l l l}
\tabletypesize{\scriptsize}
\tablecaption{\label{tab:M31RSGs} RSGs in M31}
\tablewidth{0pt}
\tablehead{
\colhead{$\alpha_{2000}$}
& \colhead{$\delta_{2000}$}
& \colhead{$\rho$\tablenotemark{a}}
& \colhead{$K_s$}
& \colhead{$\sigma_{K_s}$}
& \colhead{$J-K_s$}
& \colhead{$\sigma_{J-K_s}$}
& \colhead{\#obs}
& \colhead{Gaia\tablenotemark{b}}
& \colhead{Teff\tablenotemark{c,d}}
& \colhead{$\log L/L_\odot$\tablenotemark{c,e}}
& \multicolumn{3}{c}{LGGS} 
&\colhead{RV}
&\colhead{Exp.\ RV}
\\ \cline{12-14}
&&&&&&&&&&&\colhead{ID} & \colhead{$V$} & \colhead{Type}
&\colhead{km s$^{-1}$}
&\colhead{km s$^{-1}$}
}
\startdata
00 46 19.482&+41 59 49.66&0.71&16.819& 0.012& 0.901& 0.021& 2&2& 4250&4.02& \nodata                  &\nodata&\nodata   &\nodata&\nodata\\
00 46 20.759&+42 10 01.75&0.71&14.989& 0.024& 1.211& 0.025& 2&1& 3700&4.58&J004620.78+421001.8& 19.99&\nodata   &\nodata&\nodata\\
00 46 20.911&+41 55 45.69&0.75&14.785& 0.010& 1.156& 0.013& 3&1& 3800&4.69&J004620.92+415545.8& 19.76&M1 I      &  -91.3&  -90.6\\
00 46 23.582&+42 10 12.08&0.72&16.573& 0.017& 1.071& 0.018& 2&1& 3950&4.03&J004623.60+421012.2& 20.74&\nodata   &\nodata&\nodata\\
00 46 23.739&+42 01 41.33&0.72&14.722& 0.015& 1.144& 0.027& 2&1& 3800&4.73&J004623.75+420141.4& 20.00&M2 I      &  -78.2&  -66.6\\
00 46 25.587&+41 59 11.56&0.74&16.354& 0.007& 1.096& 0.009& 2&1& 3900&4.10&J004625.59+415911.6& 20.67&\nodata   &\nodata&\nodata\\
00 46 25.661&+42 13 21.76&0.74&15.280& 0.046& 1.178& 0.046& 2&1& 3750&4.48&J004625.67+421322.0& 20.14&\nodata   &\nodata&\nodata\\
00 46 25.670&+42 00 46.90&0.73&16.388& 0.008& 1.111& 0.030& 2&1& 3850&4.08&J004625.69+420046.9& 20.72&\nodata   &\nodata&\nodata\\
\enddata
\tablecomments{Table 2 is published in its entirety in the machine-readable format. A portion is shown here for guidance regarding its form and content.}
\tablenotetext{a}{Galactocentric distance.  Assumes an R25 radius of 95.3 arcmin, inclination
77.0 deg, and a position angle of the major axis of 35.0 deg.  At
a distance of 760 kpc, a $\rho$ of 1.00 corresponds to 21.07 kpc.}
\tablenotetext{b}{Gaia flag: 0=probable member, 1=uncertain membership; 2=no Gaia data.}
\tablenotetext{c}{Computed using $A_V=1.0$~mags.}
\tablenotetext{d}{Typical uncertainty 150~K.}
\tablenotetext{e}{Typical uncertainty 0.05~dex.}
\end{deluxetable}

\rotate
\begin{deluxetable}{l l l l l l l l l l l l l l l l l l}
\tabletypesize{\scriptsize}
\tablecaption{\label{tab:M33RSGs} RSGs in M33}
\tablewidth{0pt}
\tablehead{
\colhead{$\alpha_{2000}$}
& \colhead{$\delta_{2000}$}
& \colhead{$\rho$\tablenotemark{a}}
& \colhead{$K_s$}
& \colhead{$\sigma_{K_s}$}
& \colhead{$J-K_s$}
& \colhead{$\sigma_{J-K_s}$}
& \colhead{\#obs}
& \colhead{Gaia\tablenotemark{b}}
& \colhead{Teff\tablenotemark{c,d}}
& \colhead{$\log L/L_\odot$\tablenotemark{c,e}}
& \multicolumn{3}{c}{LGGS} 
&\colhead{RV}
&\colhead{Exp.\ RV}
\\ \cline{12-14}
&&&&&&&&&&&\colhead{ID} & \colhead{$V$} & \colhead{Type}
&\colhead{km s$^{-1}$}
&\colhead{km s$^{-1}$}
}
\startdata
01 32 57.73 &+30 12 25.7 &0.96&16.757& 0.013& 0.899& 0.021& 8&1& 4250&4.11&J013257.77+301226.0& 20.83&\nodata   &\nodata&\nodata\\
01 32 57.73 &+30 43 22.9 &0.70&15.688& 0.008& 0.983& 0.013& 4&1& 4100&4.49&J013257.79+304323.1& 19.49&\nodata   &\nodata&\nodata\\
01 32 57.81 &+30 35 54.7 &0.59&14.453& 0.002& 1.110& 0.004& 8&1& 3850&4.92&J013257.86+303555.0& 19.07&RSG       &\nodata&\nodata\\
01 32 57.86 &+30 24 59.9 &0.65&15.604& 0.005& 1.166& 0.009& 8&1& 3750&4.43&J013257.92+302500.1& 20.13&\nodata   &\nodata&\nodata\\
01 32 58.13 &+30 36 06.1 &0.58&13.604& 0.001& 1.059& 0.002& 8&1& 3950&5.29&J013258.18+303606.3& 18.35&M2.5I     &\nodata&\nodata\\
01 32 58.14 &+30 26 33.6 &0.62&16.644& 0.012& 0.939& 0.020& 8&1& 4150&4.13& \nodata                  &\nodata&\nodata   &\nodata&\nodata\\
01 32 58.19 &+30 40 05.8 &0.63&16.998& 0.014& 0.825& 0.021&12&1& 4350&4.05&J013258.25+304006.0& 21.60&\nodata   &\nodata&\nodata\\
01 32 58.57 &+30 53 53.3 &0.97&16.825& 0.026& 1.007& 0.040& 2&2& 4050&4.03&  \nodata                 &\nodata&\nodata   &\nodata&\nodata\\
\enddata
\tablecomments{Table 2 is published in its entirety in the machine-readable format. A portion is shown here for guidance regarding its form and content.}
\tablenotetext{a}{Galactocentric distance.  Assumes an R25 radius of 30.8 arcmin, inclination 56.0 deg, and a position angle of the major axis of 23.0 deg.  At a distance of 830 kpc, a $\rho$ of 1.00 corresponds to 7.44 kpc.}
\tablenotetext{b}{Gaia flag: 0=probable member, 1=uncertain membership; 2=no Gaia data.}
\tablenotetext{c}{Computed using $A_V=0.8$~mag}
\tablenotetext{d}{Typical uncertainty 150~K.}
\tablenotetext{e}{Typical uncertainty 0.05~dex.}
\end{deluxetable}

\begin{deluxetable}{c c c r r r r c c c l r l l}
\rotate
\tabletypesize{\scriptsize}
\tablecaption{\label{tab:SMCRSGs} RSGs in the SMC}
\tablewidth{0pt}
\tablehead{
&&&&&&&&&&&\multicolumn{2}{c}{Spectral Type} \\  \cline {11-12}
\colhead{2MASS}
&\colhead{$\alpha_{\rm J2000}$}
&\colhead{$\delta_{\rm J2000}$}
&\colhead{$K_s$}
&\colhead{$\sigma_{Ks}$}
&\colhead{$J-K_s$}
&\colhead{$\sigma_{J-Ks}$}
&\colhead{Gaia\tablenotemark{a}}
&\colhead{$T_{\rm eff}$\tablenotemark{b,c}}
&\colhead{$\log L/L_\odot$\tablenotemark{b,d}}
&\colhead{Type} &\colhead{Ref.}
&\colhead{Other ID}
&\colhead{Comments}
}
\startdata
J00450138-7350513 &00 45 01.383 &-73 50 51.34 &11.170 &0.019 &0.918 &0.032 &0&4100 &4.01 &\nodata &\nodata&\nodata&\nodata\\
J00450313-7255156 &00 45 03.135 &-72 55 15.61 &10.314 &0.023 &0.935 & 0.033 &0&4100 &4.35& G8 Ib &     5 &[GDN2015] SMC079&\nodata\\
J00450456-7305276 &00 45 04.566 &-73 05 27.69  &8.071 &0.023 &1.137 &0.033 &0& 3750 &5.14 &M2 I & 1& [M2002] SMC 005092 &RV var Ref 9\\
J00450482-7340132 &00 45 04.824 &-73 40 13.28 &12.209 &0.027 &0.791 &0.036 &0&4350 &3.66&\nodata&\nodata&\nodata&\nodata\\
J00450746-7327417 &00 45 07.464 &-73 27 41.79 &10.300 &0.025 &1.008 &0.035 &0&3950 &4.31&\nodata&\nodata&\nodata&\nodata\\
J00450758-7358102 &00 45 07.581& -73 58 10.21 &10.868 &0.025 &0.923 &0.035 &0&4100 &4.18 &G7 Iab-Ib & 5 &[GDN2015] SMC080&\nodata\\
J00450816-7325382 &00 45 08.165 &-73 25 38.22 &12.131 &0.029 &0.849 &0.038 &0&4250 &3.66&\nodata&\nodata&\nodata&\nodata\\
J00450831-7254146 &00 45 08.312 &-72 54 14.64 &12.273 &0.032 &0.920 &0.042 &0&4100 &3.57&\nodata&\nodata&\nodata&\nodata\\
\enddata
\tablecomments{Table~\ref{tab:SMCRSGs} is published in its entirety in the machine-readable format.  A portion is shown here for guidance regarding its form and content.}
\tablenotetext{a}{Membership flag based on Gaia information: 0=Probable member; 1=Uncertain member; 2=Incomplete Gaia data.}
\tablenotetext{b}{Computed assuming Av=0.4 mag.}
\tablenotetext{c}{Typical uncertainties 150~K.}
\tablenotetext{d}{Typical uncertainties 0.05~dex.}
\tablerefs{1--\citet{EmilyMC};
2--\citet{HV11423};
3--\citet{EmilyVariables};
4--\citet{TZO};
5--\citet{Dorda2018};
6--\citet{RSGWRs};
7--\citet{LMCBins};
8--\citet{NeugentRSGBinII};
9--\citet{DordaRV};
10--\citet{N330};
11--\citet{1980ApJ...242L..13E};
12--\citet{2018MNRAS.479.3101B};
13--\citet{1977AAS...30..261A};
14--\citet{1989AJ.....98..825S}.
}
\end{deluxetable}
\clearpage

\begin{deluxetable}{c c c r r r r c c c l r l l}
\rotate
\tabletypesize{\scriptsize}
\tablecaption{\label{tab:LMCRSGs} RSGs in the LMC}
\tablewidth{0pt}
\tablehead{
&&&&&&&&&&&\multicolumn{2}{c}{Spectral Type} \\  \cline {11-12}
\colhead{2MASS}
&\colhead{$\alpha_{\rm J2000}$}
&\colhead{$\delta_{\rm J2000}$}
&\colhead{$K_s$}
&\colhead{$\sigma_{Ks}$}
&\colhead{$J-K_s$}
&\colhead{$\sigma_{J-Ks}$}
&\colhead{Gaia\tablenotemark{a}}
&\colhead{$T_{\rm eff}$\tablenotemark{b,c}}
&\colhead{$\log L/L_\odot$\tablenotemark{b,d}}
&\colhead{Type} &\colhead{Ref.}
&\colhead{Other ID}
}
\startdata
05055608-7032372 &05 05 56.082& -70 32 37.29  &9.995 &0.023 &0.997& 0.031 & 0 &4050 &4.31\\ 
05055667-7035238 &05 05 56.678 &-70 35 23.87  &8.114 &0.026 &1.120 &0.033&  0 &3800 &5.00 &M4.5I       & 8  &[M2002] LMC 67982\\
05055720-6858112 &05 05 57.206& -68 58 11.23 &11.554 &0.023 &0.944& 0.035 & 0 &4150 &3.71\\ 
05055891-6830314 &05 05 58.919& -68 30 31.47 &11.599 &0.023 &0.943 &0.033 & 0 &4150 &3.70 \\
05055895-7029140 &05 05 58.956 &-70 29 14.04  &8.472 &0.023 &1.097 &0.031 & 0 &3850 &4.87 &M1I          &8 & [M2002] LMC 68098\\
05055950-7048112 &05 05 59.506 &-70 48 11.24  &8.315 &0.023 &1.168 &0.033 & 0 &3750 &4.89 &M4I          &8  &[M2002] LMC 68125\\
05060067-6916283 &05 06 00.678 &-69 16 28.38 &11.512 &0.023 &0.907& 0.031 & 0 &4200& 3.75 \\
05060114-6832559 &05 06 01.142 &-68 32 55.94 &11.839 &0.026 &0.936 &0.035 & 0 &4150 &3.60\\
\enddata
\tablecomments{Table~\ref{tab:LMCRSGs} is published in its entirety in the machine-readable format.  A portion is shown here for guidance regarding its form and content.}
\tablenotetext{a}{Membership flag based on Gaia information: 0=Probable member; 1=Uncertain member; 2=Incomplete Gaia data.}
\tablenotetext{b}{Computed assuming Av=0.7 mag.}
\tablenotetext{c}{Typical uncertainties 150~K.}
\tablenotetext{d}{Typical uncertainties 0.05~dex.}
\tablerefs{
    1--\citet{vanLoonMdot};
    2--\citet{1990AJ.....99..784H};
    3--\citet{1990AandA...236L..21H};
    4--\citet{1992AandAS...93..495M};
    5--\citet{1992AJ....103.1205P};
    6--\citet{1976AandAS...24...35S};
    7--\citet{1996ApJ...465..231O};
    8--\citet{MasseyOlsen};
    9--\citet{EmilyMC};
   10--\citet{1991ApJ...373...80C};
   11--\citet{2006AJ....132..866R};
   12--\citet{1987AandA...186..182W};
   13--\citet{1990ApJ...348...98R};
   14--\citet{1991AJ....101.1304H};
   15--\citet{1985ApJ...299..236R};
   16--\citet{Waterhouse};
   17--\citet{Dorda2018};
   18--\citet{1978AandAS...31..243R};
   19--\citet{2012MNRAS.425..355R};
   20--\citet{1979MNRAS.186..831F};
   21--\citet{1974ApJ...190L.133H};
   22--\citet{1989AJ.....97..107M};
   23--\citet{1977PWSO...2..105S};
   24--\citet{1981AAS...43..267W}.
}
\end{deluxetable}

\begin{deluxetable}{l r r r r r r}
\tablecaption{\label{tab:Fractions} Percentage of High Luminosity RSGs Observed and Predicted}
\tabletypesize{\small}
\tablewidth{0pt}
\tablehead{
&\multicolumn{6}{c}{Percentage of $\log L/L_\odot>5.0$ RSGs w.r.t.} \\ \cline{2-7} 
&\colhead{$\log L/L_\odot$}
&\colhead{$\log L/L_\odot$} 
&\colhead{$\log L/L_\odot$} 
&\colhead{$\log L/L_\odot$} 
&\colhead{$\log L/L_\odot$} 
&\colhead{$\log L/L_\odot$} \\
Model/Sample& \colhead{$\geq 4.0$}
& \colhead{$\geq 4.1$}
& \colhead{$\geq 4.2$}
& \colhead{$\geq 4.3$}
& \colhead{$\geq 4.4$}
& \colhead{$\geq 4.5$} 
}
\startdata
\multicolumn{7}{c}{Model Predictions} \\ \hline
Low $\dot{M}$ w/ Rot & 9.3 & 12.0 &16.0 & 20.3 & 24.8 &29.5 \\
Low $\dot{M}$ no Rot & 10.1 & 11.6 & 13.4 & 15.3 & 18.7 & 22.9 \\
Standard $\dot{M}$ w/ Rot & 2.8 & 3.3 & 4.1 & 5.1 & 6.1 & 7.8 \\
Standard $\dot{M}$ no Rot & 4.7 & 6.1 & 7.8 & 10.7 &13.4 &16.7 \\
10$\times$ $\dot{M}$ w/Rot & 1.3 & 1.8 & 2.6 & 3.5 & 4.7 & 6.5 \\
10$\times$ $\dot{M}$ no Rot & 0.6 & 0.9 & 1.2 & 1.5 & 2.1 & 2.9 \\
25$\times$ $\dot{M}$ w/Rot & 0.2 & 0.3 & 0.6 & 0.8 & 1.2 & 1.9 \\
25$\times$ $\dot{M}$ no Rot& 0.2 & 0.4 & 0.6 & 0.8 & 1.2 & 1.7 \\ \hline
\multicolumn{7}{c}{Observed} \\ \hline
M31 (total) & 1.7 & 3.1 & 5.3 & 7.8 & 10.7 & 14.2\\
M31 (low ext.) & 2.5 & 3.9 & 5.7 & 7.8 & 10.4 & 13.5 \\
M33 ($\rho<0.25$) & 5.8 &7.4 & 9.5 & 12.0 & 14.8 & 18.3 \\ \hline
\enddata
\end{deluxetable}

\begin{deluxetable}{l l c c c c c c}
\tablecaption{\label{tab:M31M33MostLums} Most Luminous RSGs in M31 and M33}
\tablewidth{0pt}
\tablehead{
\colhead{LGGS}
&\colhead{Sp.Type}
&\colhead{$\rho$}
&\colhead{$V$}
&\colhead{$K_s$}
&\colhead{$J-K_s$}
&\colhead{$T_{\rm eff}$}
&\colhead{$\log L/L_\odot$} \\
&&&\colhead{mag}&\colhead{mag}&\colhead{mag}
&\colhead{K}
&
}
\startdata
\multicolumn{8}{c}{M31 $Z\sim1.5$}\\ \hline
J003951.33+405303.7 &  \nodata & 0.75 & 17.84 & 12.84 &  1.18 &    3750 &   5.46\\
J004428.48+415130.9 & M1 I     & 0.47 & 18.27 & 13.06 &  1.16 &    3800 &   5.38\\
J004125.23+411208.9 & M0 I     & 0.48 & 19.07 & 12.93 &  1.30 &    3500 &   5.36\\
J004514.91+413735.0 & M1 I     & 0.61 & 19.55 & 12.89 &  1.34 &    3450 &   5.35\\
J004503.35+413026.3 &  \nodata & 0.67 & 18.82 & 13.43 &  0.99 &   4100 &   5.33\\
J004312.43+413747.1 & M2 I     & 0.44 & 19.36 & 12.97 &  1.35 &   3450 &   5.32\\
J004047.22+404445.5 & M2 I     & 0.40 & 19.33 & 13.10 &  1.26 &    3600 &   5.31\\
J004059.50+404542.6 & M0 I     & 0.38 & 18.71 & 13.23 &  1.18 &    3750 &   5.31\\
J004138.35+412320.7 &  \nodata & 0.67 & 18.47 & 13.39 &  1.07 &    3950 &   5.30\\
J004148.74+410843.0 & M2 I     & 0.24 & 19.54 & 13.36 &  1.13 &    3850 &   5.28\\ \hline
\multicolumn{8}{c}{M33 Inner $Z\sim1.0$}\\ \hline
J013336.64+303532.3 & RSG      & 0.18 & 18.27 & 13.18 &  0.99 &   4100 &   5.49\\
J013352.96+303816.0 &  \nodata & 0.06 & 19.42 & 13.18 &  1.08 &   3900 &   5.45\\
J013357.08+303817.8 &  \nodata & 0.10 & 18.86 & 13.61 &  1.01 &   4050 &   5.31\\
J013335.90+303344.5 &  \nodata & 0.22 & 19.35 & 13.37 &  1.24 &   3650 &   5.29\\
J013351.47+303640.3 &  \nodata & 0.11 & 19.92 & 13.47 &  1.17 &   3750 &   5.28\\
J013350.84+304403.1 & RSG      & 0.17 & 19.00 & 13.56 &  1.14 &   3800 &   5.26\\
J013403.73+304202.4 & RSG      & 0.14 & 19.43 & 13.60 &  1.19 &   3750 &   5.22\\
J013344.10+304425.1 &  \nodata & 0.23 & 19.15 & 13.58 &  1.23 &   3650 &   5.21\\
J013409.63+303907.6 &  \nodata & 0.23 & 20.13 & 13.36 &  1.41 &   3350 &   5.20\\
J013338.77+303532.9 &  \nodata & 0.16 & 19.52 & 13.77 &  1.13 &   3850 &   5.19\\
\hline
\multicolumn{8}{c}{M33 Middle $Z\sim0.5$}\\ \hline
J013414.27+303417.7 & RSG      & 0.40 & 18.34 & 12.78 &  1.25 &   3600 &   5.52\\
J013350.62+303230.3 & RSG      & 0.27 & 18.12 & 12.98 &  1.15 &   3800 &   5.49\\
J013358.54+303419.9 & OB/RSG? & 0.25 & 17.90 & 13.21 &  1.01 &   4050 &   5.47\\
J013412.27+305314.1 & RSG      & 0.47 & 18.48 & 13.37 &  1.12 &   3850 &   5.35\\
J013328.17+304741.5 &  \nodata & 0.48 & 18.27 & 13.35 &  1.12 &   3850 &   5.35\\
J013312.35+303033.9 &  \nodata & 0.45 & 18.42 & 13.51 &  1.12 &   3850 &   5.29\\
J013309.10+303017.8 & RSG      & 0.48 & 18.06 & 13.74 &  0.98 &   4100 &   5.28\\
J013321.44+304045.4 & RSG      & 0.37 & 19.17 & 13.62 &  1.07 &   3950 &   5.28\\
J013319.13+303642.5 & M3 I     & 0.35 & 18.72 & 13.57 &  1.13 &   3850 &   5.27\\
J013314.31+302952.9 & RSG     & 0.44 & 19.69 & 13.47 &  1.22 &   3650 &   5.26\\
\hline
\multicolumn{8}{c}{M33 Outer $Z\sim0.3$}\\ \hline
J013241.94+302047.5 &  \nodata & 0.84 & 17.67 & 13.29 &  1.06 &   3950 &   5.41\\
J013307.60+304259.0 &  \nodata & 0.57 & 18.35 & 13.41 &  1.11 &   3850 &   5.34\\
J013430.75+303218.8 &  \nodata & 0.63 & 19.21 & 13.38 &  1.16 &   3800 &   5.33\\
J013303.54+303201.2 & RSG      & 0.52 & 18.88 & 13.46 &  1.11 &   3850 &   5.32\\
J013233.77+302718.8 & RSG      & 0.84 & 18.32 & 13.43 &  1.17 &   3750 &   5.30\\
J013305.48+303138.5 &  \nodata & 0.50 & 18.55 & 13.44 &  1.17 &   3750 &   5.29\\
J013423.29+305655.0 & RSG      & 0.61 & 18.96 & 13.51 &  1.12 &   3850 &   5.29\\
J013258.18+303606.3 & M2.5 I    & 0.58 & 18.35 & 13.60 &  1.06 &   3950 &   5.29\\
J013329.47+301848.3 &  \nodata & 0.72 & 18.75 & 13.55 &  1.12 &   3850 &   5.28\\
J013312.38+302453.2 & RSG       & 0.56 &18.29 & 13.66 &  1.06  &   3950 &  5.26 \\
\enddata
\end{deluxetable}

\begin{deluxetable}{l l r r r c c l}
\tabletypesize{\scriptsize}
\tablecaption{\label{tab:MCsMostLums}Most Luminous RSGs in the Magellanic Clouds}
\tablewidth{0pt}
\tablehead{
\colhead{2MASS}
&\colhead{Sp.\ Type\tablenotemark{a}}
&\colhead{$V$\tablenotemark{b}}
&\colhead{$K_s$}
&\colhead{$J-K_s$}
&\colhead{$T_{\rm eff}$}
&\colhead{$\log L/L_\odot$}
&\colhead{Other ID}
}
\startdata
\multicolumn{7}{c}{LMC $Z\sim0.5$} \\ \hline
J05265311-6850002& K/M  & 12.24&6.62&1.13&3800&5.59& HD 269551. Close OB visual companion.\\
J05080770-6548486& \nodata & \nodata & 6.63&1.15&3750&5.58& WOH S170   \\
J05294221-6857173& M2 & \nodata &6.89&1.04&3950&5.53& [M2002] LMC 145013 \\
J05041413-6716143& K/M & 11.86 & 6.78&1.23&3650&5.48& SV* HV 888\\
J04525366-6655520& K5-M3?\tablenotemark{c} & 11.14 & 7.09&0.99&4050&5.47& SP77 21-12\\
J04443612-7043022& \nodata &13.80 & 6.78&1.27&3550&5.46& SV* HV 12463\\
J05070565-7032439& M1~I & 11.76 & 7.04&1.08&3900&5.45& SV* HV 5618\\
J04553485-6926556&\nodata& 11.76& 7.11&1.12&3800&5.40&[M2002] LMC 25320 7\arcsec visual double with OB\\
J05121313-6804555& M0.5~I &11.62 &  7.32&0.97&4100&5.39&SV* HV 2362\\
J05352832-6656024& M3~I &12.11 &  7.26&1.10&3850&5.35& [W60] D44\\
J05264293-6724326& K/M &11.94 & 7.38&1.02&4000&5.34& 	[M2002] LMC 136404 \\
J05302094-6720054& M4~I &12.79 &  7.45&1.00&4050&5.33&[M2002] LMC 147199 \\
J05355522-6909594& \nodata &12.87 & 7.14&1.41&3300&5.24&	OGLE BRIGHT-LMC-LPV-52 \\
J04550307-6929127& \nodata &14.44 & 7.20&1.46&3200&5.19& 	[M2002] LMC 23095 \\ 
J05315335-6640432 & M3/4~I  & 12.31  & 7.57&1.18 &3700 &5.18 &    TRM 87\\  \hline
\multicolumn{7}{c}{SMC $Z\sim0.2$} \\ \hline
J00510385-7243176&M3 Iab    &11.39 & 7.43&0.96&4050&5.49&[M2002] SMC 018592\\
J00482702-7312123&M0 I      &12.20 & 7.77&1.00&4000&5.33&[M2002] SMC 010889\\
J01032764-7252096&K2 Iab  & 11.85  & 7.78&1.02&3950&5.32&[M2002] SMC 056389\\
J01291844-7301590&K:I+B5:I & 11.53       & 7.95&0.91&4150&5.30&[M2002] SMC 83202                  \\
J01093824-7320024&K5-M0/M4~I\tablenotemark{c}&11.74 & 7.79&1.06&3900&5.29&[M2002] SMC 069886\\
J00505609-7215060&M0 I   &11.98  & 7.85&1.05&3900&5.27&[M2002] SMC 018136\\
J01004151-7210371&K5-M0 I &12.17   & 7.96&0.99&4000&5.26&[M2002] SMC 049478\\
J01253885-7321552&K/M &11.84        & 7.98&1.05&3900&5.22& [M2002] SMC 81961             \\
J01005519-7137529&K0/M4-5 I & 11.81 & 8.04&1.03&3900&5.20&SV* HV 11423      \\
J01043820-7201270&K2-3 I    & 11.98 & 8.10&1.00&4000&5.20&[M2002] SMC 059803\\
J01064766-7216118&K3 Ia     & 11.87 & 8.31&0.93&4100&5.15&[M2002] SMC 064663\\
J00450456-7305276&M2 I      & 12.90 & 8.07&1.14&3750&5.14&[M2002] SMC 005092\\
J00512967-7310442&M0 I      & 12.33 & 8.15&1.07&3850&5.14&[M2002] SMC 020133\\
J00534794-7202095&M2 Ia-Iab & 12.43 & 8.25 &1.02 & 3950 & 5.13&[S84d] 105-11  \\
J00490034-7259358&K5-M0 I & 12.44 & 8.28 & 1.00 & 4000 & 5.13 & [M2002] SMC 12322\\ 
\enddata
\tablecomments{As discussed in the text, extremely dusty RSGs, such as WOH G64 \citep{1986ApJ...302..675E} and the recently discovered ``LMC3" \citep{2022arXiv220911239D}, will be missing
from these lists, as their $J-K$ colors were so red as to remove them from our RSG lists.}
\tablenotetext{a}{References for spectral types are given in Table~\ref{tab:SMCRSGs} and \ref{tab:LMCRSGs}.}
\tablenotetext{b}{$V$ band photometry from \citet{ZaritskyLMC} for the LMC and from \citet{Massey2002} for the SMC.}
\tablenotetext{c}{Spectral variable?}
\end{deluxetable}
\end{document}